\pdfoutput=1
\documentclass[12pt,a4paper]{article}
\usepackage{ifthen} % for conditional statements
\newboolean{pdflatex}
\setboolean{pdflatex}{true} % False for eps figures 

\newboolean{articletitles}
\setboolean{articletitles}{true} % False removes titles in references

\newboolean{uprightparticles}
\setboolean{uprightparticles}{false} %True for upright particle symbols

\def\paperauthors{LHCb collaboration} % Leave as is for PAPER, CONF and FIGURE

\def\paperasciititle{Updated measurement of CP violation and polarisation in Bs0 -> J/psi K*~(892)0 decays} % Set ASCII title here !! MAKE sure it's only ASCII characters !! 
\def\papertitle{
Updated measurement of 
$C\!P$
violation and polarisation in 
\mbox{$B^0_s  \rightarrow  J\mskip -3mu/\mskip -2mu\psi  \Kbar{}^{*}\kern-1pt(892)^{0}$}
decays
}

\def\paperkeywords{{High Energy Physics}, {LHCb}} % Comma separated list
\def\papercopyright{\the\year\ CERN for the benefit of the LHCb collaboration} % new since 9/Apr/2018
\def\paperlicence{CC BY 4.0 licence}
\def\paperlicenceurl{https://creativecommons.org/licenses/by/4.0/}

\newif\ifEnableSectionTOCLinks
\EnableSectionTOCLinksfalse % deactivated

\usepackage[top=1in, bottom=1.25in, left=1in, right=1in]{geometry}

\usepackage{microtype}
\usepackage{lineno}  % for line numbering during review
\usepackage{xspace} % To avoid problems with missing or double spaces after
\usepackage{caption} %these three command get the figure and table captions automatically small

\usepackage{graphicx}  % to include figures (can also use other packages)
\usepackage{color}
\usepackage{colortbl}
\graphicspath{{./figs/}} % Make Latex search fig subdir for figures
\usepackage{amsmath} % Adds a large collection of math symbols
\usepackage{amssymb}
\usepackage{amsfonts}
\usepackage{upgreek} % Adds in support for greek letters in roman typeset

\newcommand*\patchAmsMathEnvironmentForLineno[1]{%
\expandafter\let\csname old#1\expandafter\endcsname\csname #1\endcsname
\expandafter\let\csname oldend#1\expandafter\endcsname\csname
end#1\endcsname
 \renewenvironment{#1}%
   {\linenomath\csname old#1\endcsname}%
   {\csname oldend#1\endcsname\endlinenomath}%
}
\newcommand*\patchBothAmsMathEnvironmentsForLineno[1]{%
  \patchAmsMathEnvironmentForLineno{#1}%
  \patchAmsMathEnvironmentForLineno{#1*}%
}
\AtBeginDocument{%
\patchBothAmsMathEnvironmentsForLineno{equation}%
\patchBothAmsMathEnvironmentsForLineno{align}%
\patchBothAmsMathEnvironmentsForLineno{flalign}%
\patchBothAmsMathEnvironmentsForLineno{alignat}%
\patchBothAmsMathEnvironmentsForLineno{gather}%
\patchBothAmsMathEnvironmentsForLineno{multline}%
\patchBothAmsMathEnvironmentsForLineno{eqnarray}%
}

\usepackage[pdftex,
            pdfauthor={\paperauthors},
            pdftitle={\paperasciititle},
            pdfkeywords={\paperkeywords}]{hyperref}
\usepackage{hyperxmp}
\hypersetup{
    pdfcopyright={Copyright (C) \papercopyright},
    pdflicenseurl={\paperlicenceurl}
}
\usepackage[colorinlistoftodos,textsize=scriptsize]{todonotes}

\usepackage[bottom,flushmargin,hang,multiple]{footmisc}

\usepackage[all]{hypcap} % Internal hyperlinks to floats.

\usepackage{xspace}
\usepackage{upgreek}

\def\lhcb   {\mbox{LHCb}\xspace}

\def\belle  {\mbox{Belle}\xspace}

\def\MagUp {\mbox{\em Mag\kern -0.05em Up}\xspace}

\ifthenelse{\boolean{uprightparticles}}%
{

 \def\Pmu         {\ensuremath{\upmu}\xspace}

 \def\Ppi         {\ensuremath{\uppi}\xspace}
 
 \def\Prho        {\ensuremath{\uprho}\xspace}

 \def\Ppsi        {\ensuremath{\uppsi}\xspace}

 \def\PDelta      {\ensuremath{\Delta}\xspace}
 \def\PXi         {\ensuremath{\Xi}\xspace}
 \def\PLambda     {\ensuremath{\Lambda}\xspace}
 \def\PSigma      {\ensuremath{\Sigma}\xspace}
 \def\POmega      {\ensuremath{\Omega}\xspace}
 \def\PUpsilon    {\ensuremath{\Upsilon}\xspace}
 \let\oldPi\Pi
 \def\PPi         {\ensuremath{\oldPi}\xspace}

 \def\PB      {\ensuremath{\mathrm{B}}\xspace}
 \def\PD      {\ensuremath{\mathrm{D}}\xspace}
 \def\PJ      {\ensuremath{\mathrm{J}}\xspace}
 \def\PK      {\ensuremath{\mathrm{K}}\xspace}
 \def\Pb      {\ensuremath{\mathrm{b}}\xspace}
 \def\Pc      {\ensuremath{\mathrm{c}}\xspace}
 \def\Pd      {\ensuremath{\mathrm{d}}\xspace}

 \def\Pp      {\ensuremath{\mathrm{p}}\xspace}

 \def\Ps      {\ensuremath{\mathrm{s}}\xspace}

 \def\thebaroffset{0.0em}
}
{

 \def\Pmu         {\ensuremath{\mu}\xspace}

 \def\Ppi         {\ensuremath{\pi}\xspace}
 
 \def\Prho        {\ensuremath{\rho}\xspace}

 \def\Ppsi        {\ensuremath{\psi}\xspace}
 
 \mathchardef\PDelta="7101
 \mathchardef\PXi="7104
 \mathchardef\PLambda="7103
 \mathchardef\PSigma="7106
 \mathchardef\POmega="710A
 \mathchardef\PUpsilon="7107
 \mathchardef\PPi="7105
 \def\PB      {\ensuremath{B}\xspace}
 \def\PD      {\ensuremath{D}\xspace}
 \def\PJ      {\ensuremath{J}\xspace}
 \def\PK      {\ensuremath{K}\xspace}
 \def\Pb      {\ensuremath{b}\xspace}
 \def\Pc      {\ensuremath{c}\xspace}
 \def\Pd      {\ensuremath{d}\xspace}

 \def\Pp      {\ensuremath{p}\xspace}

 \def\Ps      {\ensuremath{s}\xspace}

 \def\thebaroffset{0.18em}
}
\newcommand{\offsetoverline}[2][\thebaroffset]{\kern #1\overline{\kern -#1 #2}}%

\makeatletter
\ifcase \@ptsize \relax% 10pt
  \newcommand{\miniscule}{\@setfontsize\miniscule{4}{5}}% \tiny: 5/6
\or% 11pt
  \newcommand{\miniscule}{\@setfontsize\miniscule{5}{6}}% \tiny: 6/7
\or% 12pt
  \newcommand{\miniscule}{\@setfontsize\miniscule{5}{6}}% \tiny: 6/7
\fi
\makeatother

\DeclareRobustCommand{\optbar}[1]{\shortstack{{\miniscule (\rule[.5ex]{1.25em}{.18mm})}
  \\ [-.7ex] $#1$}}

\def\mup        {{\ensuremath{\Pmu^+}}\xspace}
\def\mun        {{\ensuremath{\Pmu^-}}\xspace} % muon negative (\mum is taken)

\def\dquark    {{\ensuremath{\Pd}}\xspace}

\def\squark    {{\ensuremath{\Ps}}\xspace}

\def\cquark    {{\ensuremath{\Pc}}\xspace}
\def\cquarkbar {{\ensuremath{\overline \cquark}}\xspace}
\def\ccbar     {{\ensuremath{\cquark\cquarkbar}}\xspace}
\def\bquark    {{\ensuremath{\Pb}}\xspace}
\def\bquarkbar {{\ensuremath{\overline \bquark}}\xspace}

\def\pion   {{\ensuremath{\Ppi}}\xspace}

\def\pip    {{\ensuremath{\pion^+}}\xspace}
\def\pim    {{\ensuremath{\pion^-}}\xspace}

\def\rhomeson {{\ensuremath{\Prho}}\xspace}
\def\rhoz     {{\ensuremath{\rhomeson^0}}\xspace}

\def\kaon    {{\ensuremath{\PK}}\xspace}
\def\Kbar    {{\ensuremath{\offsetoverline{\PK}}}\xspace}

\def\KorKbar {\kern \thebaroffset\optbar{\kern -\thebaroffset \PK}{}\xspace}

\def\Kp      {{\ensuremath{\kaon^+}}\xspace}
\def\Km      {{\ensuremath{\kaon^-}}\xspace}

\def\Kstarz  {{\ensuremath{\kaon^{*0}}}\xspace}
\def\Kstarzb {{\ensuremath{\Kbar{}^{*0}}}\xspace}

\def\D       {{\ensuremath{\PD}}\xspace}

\def\DorDbar {\kern \thebaroffset\optbar{\kern -\thebaroffset \PD}\xspace}

\def\Dp      {{\ensuremath{\D^+}}\xspace}
\def\Dm      {{\ensuremath{\D^-}}\xspace}

\def\DpDm    {\ensuremath{\Dp {\kern -0.16em \Dm}}\xspace}

\def\B       {{\ensuremath{\PB}}\xspace}
\def\Bbar    {{\ensuremath{\offsetoverline{\PB}}}\xspace}

\def\BorBbar {\kern \thebaroffset\optbar{\kern -\thebaroffset \PB}\xspace}

\def\Bd      {{\ensuremath{\B^0}}\xspace}
\def\Bdb     {{\ensuremath{\Bbar{}^0}}\xspace}
\def\BdorBdbar {\kern \thebaroffset\optbar{\kern -\thebaroffset \Bd}\xspace}
\def\Bu      {{\ensuremath{\B^+}}\xspace}
\def\Bub     {{\ensuremath{\B^-}}\xspace}

\def\Bs      {{\ensuremath{\B^0_\squark}}\xspace}
\def\Bsb     {{\ensuremath{\Bbar{}^0_\squark}}\xspace}
\def\BsorBsbar {\kern \thebaroffset\optbar{\kern -\thebaroffset \Bs}\xspace}

\def\jpsi     {{\ensuremath{{\PJ\mskip -3mu/\mskip -2mu\Ppsi}}}\xspace}

\def\Y#1S{\ensuremath{\PUpsilon{(#1S)}}\xspace}

\def\FourS {{\Y4S}\xspace}

\def\proton      {{\ensuremath{\Pp}}\xspace}

\def\Lz          {{\ensuremath{\PLambda}}\xspace}

\def\LorLbar     {\kern \thebaroffset\optbar{\kern -\thebaroffset \PLambda}\xspace}

\def\Lb           {{\ensuremath{\Lz^0_\bquark}}\xspace}

\def\BF         {{\ensuremath{\mathcal{B}}}\xspace}
\def\BR         {\BF}

\newcommand{\decay}[2]{\mbox{\ensuremath{#1\!\to #2}}\xspace}

\def\to                 {\ensuremath{\rightarrow}\xspace}

\def\grpsuthree {{\ensuremath{\mathrm{SU}(3)}}\xspace}

\def\CP                {{\ensuremath{C\!P}}\xspace}

\newcommand{\phis}{{\ensuremath{\phi_{\squark}}}\xspace}

\def\BsToJPsiPhi  {\decay{\Bs}{\jpsi\phi}}

\def\AT#1     {\ensuremath{A_{\mathrm{T}}^{#1}}\xspace}           % 2

\def\C#1      {\ensuremath{\mathcal{C}_{#1}}\xspace}                       % 9
\def\Cp#1     {\ensuremath{\mathcal{C}_{#1}^{'}}\xspace}                    % 7
\def\Ceff#1   {\ensuremath{\mathcal{C}_{#1}^{\mathrm{(eff)}}}\xspace}        % 9  
\def\Cpeff#1  {\ensuremath{\mathcal{C}_{#1}^{'\mathrm{(eff)}}}\xspace}       % 7
\def\Ope#1    {\ensuremath{\mathcal{O}_{#1}}\xspace}                       % 2
\def\Opep#1   {\ensuremath{\mathcal{O}_{#1}^{'}}\xspace}                    % 7

\newcommand{\nospaceunit}[1]{\ensuremath{\text{#1}}}
\newcommand{\aunit}[1]{\ensuremath{\text{\,#1}}}
\newcommand{\tev}{\aunit{Te\kern -0.1em V}\xspace}
\newcommand{\gev}{\aunit{Ge\kern -0.1em V}\xspace}
\newcommand{\mev}{\aunit{Me\kern -0.1em V}\xspace}
\newcommand{\kev}{\aunit{ke\kern -0.1em V}\xspace}
\newcommand{\ev}{\aunit{e\kern -0.1em V}\xspace}

\newcommand{\mevc}{\ensuremath{\aunit{Me\kern -0.1em V\!/}c}\xspace}
\newcommand{\gevc}{\ensuremath{\aunit{Ge\kern -0.1em V\!/}c}\xspace}
\newcommand{\mevcc}{\ensuremath{\aunit{Me\kern -0.1em V\!/}c^2}\xspace}
\newcommand{\gevcc}{\ensuremath{\aunit{Ge\kern -0.1em V\!/}c^2}\xspace}
\def\mum  {\ensuremath{\,\upmu\nospaceunit{m}}\xspace}

\def\fb   {\ensuremath{\aunit{fb}}\xspace}
\def\invfb   {\ensuremath{\fb^{-1}}\xspace}

\newcommand{\stat}{\aunit{(stat)}\xspace}
\newcommand{\syst}{\aunit{(syst)}\xspace}

\newcommand{\chisq}{\ensuremath{\chi^2}\xspace}
\newcommand{\chisqndf}{\ensuremath{\chi^2/\mathrm{ndf}}\xspace}
\newcommand{\chisqip}{\ensuremath{\chi^2_{\text{IP}}}\xspace}

\def\deriv {\ensuremath{\mathrm{d}}}

\def\gsim{{~\raise.15em\hbox{$>$}\kern-.85em
          \lower.35em\hbox{$\sim$}~}\xspace}
\def\lsim{{~\raise.15em\hbox{$<$}\kern-.85em
          \lower.35em\hbox{$\sim$}~}\xspace}

\newcommand{\Real}{\ensuremath{\mathcal{R}e}\xspace}
\newcommand{\Imag}{\ensuremath{\mathcal{I}m}\xspace}

\def\sPlot{\mbox{\em sPlot}\xspace}

\def\pt         {\ensuremath{p_{\mathrm{T}}}\xspace}

\def\ptot       {\ensuremath{p}\xspace}

\def\mrad{\aunit{mrad}\xspace}
\def\rad{\aunit{rad}\xspace}

\def\evtgen     {\mbox{\textsc{EvtGen}}\xspace}

\def\geant      {\mbox{\textsc{Geant4}}\xspace}

\def\photos     {\mbox{\textsc{Photos}}\xspace}

\def\pythia     {\mbox{\textsc{Pythia}}\xspace}

\def\tell1  {TELL1\xspace}
\def\ukl1   {UKL1\xspace}

\newcommand{\eg}{\mbox{\itshape e.g.}\xspace}
\newcommand{\ie}{\mbox{\itshape i.e.}\xspace}

\newcommand{\lhcborcid}[1]{\href{https://orcid.org/#1}{\hspace*{0.1em}\raisebox{-0.45ex}{\includegraphics[width=1em]{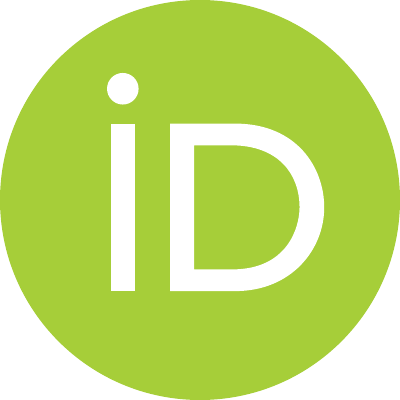}}}}

\hypersetup{
  colorlinks   = true, %Colours links instead of ugly boxes
  urlcolor     = blue, %Colour for external hyperlinks
  linkcolor    = blue, %Colour of internal links
  citecolor    = red   %Colour of citations
}

\ifEnableSectionTOCLinks
    \usepackage[explicit]{titlesec} % to change headings
    
    \let\oldcontentsline\contentsline
    \renewcommand

    \titleformat{\section}{\normalfont\Large\bf}{\hyperlink{tocsection.\thesection}{{\thesection} \parbox[t]{\dimexpr\textwidth-1pc}{#1}}}{1pc}{}

    \titleformat{\subsection}{\normalfont\bf}{\hyperlink{tocsubsection.\thesubsection}{{\thesubsection} \parbox[t]{\dimexpr\textwidth-1pc}{#1}}}{1pc}{}

    \titleformat{name=\section,numberless}[display]{}{}{0pt}{\normalfont\Huge\bfseries #1}
\fi

\usepackage{cite} % Allows for ranges in citations
\usepackage{mciteplus}
\usepackage{longtable} % only for template; not usually to be used in PAPERs

\newcommand{\KstarIz}{\ensuremath{\kaon^{*}\kern-1pt(892)^{0}}\xspace} % Alias for \Kstarz if needed, standard is K^{*0}
\newcommand{\KstarIzb}{\ensuremath{\Kbar{}^{*}\kern-1pt(892)^{0}}\xspace} % Alias for \Kstarzb if needed, standard is \Kbar^{*0}

\newcommand{\KstarIIz}{\ensuremath{\kaon^{*}_{2}\kern-1pt(1430)^{0}}\xspace} % K*_2(1430)^0
\newcommand{\KstarIIzb}{\ensuremath{\Kbar{}^{*}_{2}\kern-1pt(1430)^{0}}\xspace} % \bar{K}*_2(1430)^0
\newcommand{\Kstarzzb}{\ensuremath{\Kbar{}^{*}_{0}\kern-1pt(700)^{0}}\xspace} % K*_0(700) aka kappa - check if used/needed

\newcommand{\rhoIz}{\ensuremath{\rhomeson\kern-0.001pt(770)^{0}}\xspace}

\newcommand{\LbJpsipK}{\decay{\Lb}{\jpsi\proton\Km}}
\newcommand{\LbJpsippi}{\decay{\Lb}{\jpsi\proton\pim}}
\newcommand{\BsJpsipipi}{\decay{\Bs}{\jpsi\pip\pim}}

\newcommand{\BsJpsiKK}{\decay{\Bs}{\jpsi\Kp\Km}} % Already defined in user list, redundant?

\newcommand{\BdJpsiRhoI}{\decay{\Bd}{\jpsi \rhoIz}} % Standard B0 -> J/psi rho0(pi+pi-)

\newcommand{\swave}{\text{S-wave}\xspace}
\newcommand{\pwave}{\text{P-wave}\xspace}
\newcommand{\dwave}{\text{D-wave}\xspace}

\newcommand{\deltaPa}{\ensuremath{\delta_{\parallel}}\xspace} % P-wave phase (parallel)
\newcommand{\deltaPe}{\ensuremath{\delta_{\perp}}\xspace} % P-wave phase (perpendicular)

\newcommand{\fL}{\ensuremath{f_{0}}\xspace} % P-wave longitudinal fraction (using f_0 to match A_0)
\newcommand{\fpar}{\ensuremath{f_{\parallel}}\xspace} % P-wave parallel fraction
\newcommand{\fpa}{\ensuremath{\fpar}}

\newcommand{\ACPL}{\ensuremath{A^{\CP}_{0}}\xspace}
\newcommand{\ACPpa}{\ensuremath{A^{\CP}_{\parallel}}\xspace}
\newcommand{\ACPpe}{\ensuremath{A^{\CP}_{\perp}}\xspace}
\newcommand{\ACPS}{\ensuremath{A^{\CP}_{S}}\xspace} % Use A_{\CP}^{\text{S}} if preferred

\newcommand{\AsBinZero}{\ensuremath{F_S^{826-861}}\xspace}
\newcommand{\AsBinOne}{\ensuremath{F_S^{861-896}}\xspace}
\newcommand{\AsBinTwo}{\ensuremath{F_S^{896-931}}\xspace}
\newcommand{\AsBinThree}{\ensuremath{F_S^{931-966}}\xspace}

\newcommand{\dsBinZero}{\ensuremath{\delta_S^{826-861}}\xspace}
\newcommand{\dsBinOne}{\ensuremath{\delta_S^{861-896}}\xspace}
\newcommand{\dsBinTwo}{\ensuremath{\delta_S^{896-931}}\xspace}
\newcommand{\dsBinThree}{\ensuremath{\delta_S^{931-966}}\xspace}

\newcommand{\mkpi}{\ensuremath{m_{K\pi}}\xspace} % Mass of Kpi system

\newcommand{\thetaK}{\ensuremath{\theta_{K}}}       % Angle of K in Kpi system CM relative to Kpi direction in Bs CM
\newcommand{\costmu }{\ensuremath{\cos \theta_{\mu}}    \xspace}

\newcommand{\costk  }{\ensuremath{\cos \theta_{K}}      \xspace}

\newcommand{\cosSqtmu }{\ensuremath{\cos^{2} \theta_{\mu}}  \xspace}

\newcommand{\cosSqtk  }{\ensuremath{\cos^{2} \theta_{K}}    \xspace}

\newcommand{\thetamu  }{\ensuremath{\theta_{\mu}}\xspace}
\newcommand{\phihel   }{\ensuremath{\phi_{h}}\xspace}

\newcommand{\fdfs}{\ensuremath{\frac{f_d}{f_s}}\xspace}

\newcommand{\Kpi     }{{\ensuremath{K\pi}}\xspace}

\newcommand{\Kmpip   }{{\ensuremath{K^-\pi^+}}\xspace}

\newcommand{\BdJpsiKst}{\ensuremath{\Bd\to \jpsi K^{*0}}\xspace}

\newcommand{\BsJpsiKst}{\ensuremath{\Bs\to \jpsi \Kstarzb}\xspace}

\newcommand{\BsJpsiKstI}{\ensuremath{\Bs \to \jpsi \KstarIzb}\xspace}

\newcommand{\BdJpsiKstI}{\ensuremath{\Bd \to \jpsi \KstarIz)}\xspace}

\newcommand{\BdJpsiRho}{\ensuremath{\Bd\to \jpsi \rhoz}\xspace}

\newcommand{\BdJpsipipi}{\ensuremath{\Bd\to \jpsi \pi^+\pi^-}\xspace}
\newcommand{\BRof}[1]{\ensuremath{{\cal B}(#1)}\xspace}

\usepackage[T1]{fontenc} % Ensure proper encoding for \DH

\usepackage{rotating}
\usepackage{multirow}

\usepackage{subfig}%subfigures
\usepackage{tabularx} %tabularx
\usepackage{booktabs}

\usepackage{
    placeins
} % \FloatBarrier  % Prevents floats from moving beyond this point

\usepackage{cleveref} % clever reference --> cref
\usepackage{xfrac} % \sfrac{1}{1}

\usepackage{bm}
\usepackage{xspace}

\begin{document}

%%%%%%%%%%%%%%%%%%%%%%%%%
%%%%% Title     %%%%%%%%%
%%%%%%%%%%%%%%%%%%%%%%%%%
\renewcommand{\thefootnote}{\fnsymbol{footnote}}
\setcounter{footnote}{1}

% %%%%%%% CHOOSE TITLE PAGE--------
%\onecolumn
%\input{title-LHCb-INT}
%\input{title-LHCb-ANA}
%\input{title-LHCb-CONF}
%\input{title-LHCb-FIGURE}
% ===============================================================================
% Purpose: LHCb-PAPER journal paper title page template
% Author: 
% Created on: 2010-09-25
% ===============================================================================

%%%%%%%%%%%%%%%%%%%%%%%%%
%%%%%  TITLE PAGE  %%%%%%
%%%%%%%%%%%%%%%%%%%%%%%%%
\begin{titlepage}
\pagenumbering{roman}

% Header ---------------------------------------------------
\vspace*{-1.5cm}
\centerline{\large EUROPEAN ORGANIZATION FOR NUCLEAR RESEARCH (CERN)}
\vspace*{1.5cm}
\noindent
\begin{tabular*}{\linewidth}{lc@{\extracolsep{\fill}}r@{\extracolsep{0pt}}}
\ifthenelse{\boolean{pdflatex}}% Logo format choice
{\vspace*{-1.5cm}\mbox{\!\!\!\includegraphics[width=.14\textwidth]{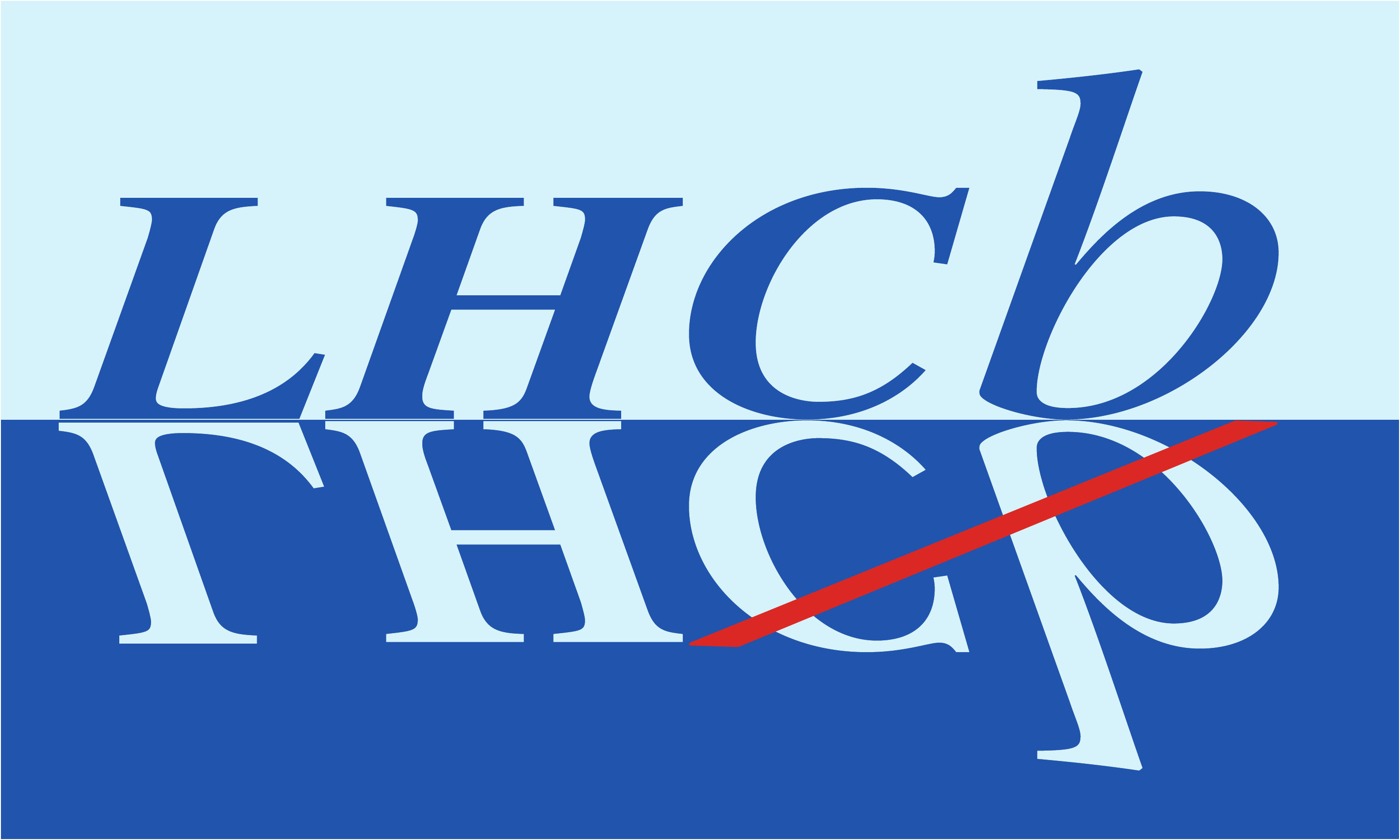}} & &}%
{\vspace*{-1.2cm}\mbox{\!\!\!\includegraphics[width=.12\textwidth]{figs/lhcb-logo.eps}} & &}%
\\
 & & CERN-EP-2025-131 \\  % ID 
 & & LHCb-PAPER-2025-020 \\  % ID 
 % & & \today \\ % Date - Can also hardwire e.g.: 23 March 2010
 & & 23 October 2025 \\ % Date - Can also hardwire e.g.: 23 March 2010
 & & \\
% not in paper \hline
\end{tabular*}

\vspace*{4.0cm}

% Title --------------------------------------------------
{\normalfont\bfseries\boldmath\huge
\begin{center}
% DO NOT EDIT HERE. Instead edit macro in main.tex to keep metadata correct
  \papertitle 
\end{center}
}

\vspace*{1.5cm}

% Authors -------------------------------------------------
\begin{center}
%In the footnote, replace 'paper' by 'Letter' in case of submission to PRL or PLB 
% Edit macro in main.tex to keep metadata correct
\paperauthors\footnote{Authors are listed at the end of this paper.}
\end{center}

\vspace{\fill}

% Abstract -----------------------------------------------
\begin{abstract}
\noindent
   A time-integrated angular analysis of the decay 
   \mbox{$B^0_s  \rightarrow  J\mskip -3mu/\mskip -2mu\psi  \Kbar{}^{*}\kern-1pt(892)^{0}$}, 
   with 
   \mbox{$J\mskip -3mu/\mskip -2mu\psi  \rightarrow  \mu^{+}  \mu^{-}$} 
   and 
   \mbox{$\Kbar{}^{*}\kern-1pt(892)^{0}  \rightarrow  K^{-}  \pi^{+}$}
   , is presented.
  The analysis employs a sample of proton-proton collision data collected by the \mbox{LHCb} experiment during 2015--2018 at a centre-of-mass energy of \mbox{$13 \text{\,Te\kern -0.1em V}$}, corresponding to an integrated luminosity of \mbox{$6 \text{\,fb}^{-1}$}.
  A simultaneous maximum-likelihood fit is performed to the angular distributions in bins of the \mbox{$K^{-}  \pi^{+}$} mass. 
  This fit yields measurements of the $C\!P$-averaged polarisation fractions and $C\!P$ asymmetries for the \pwave component of the \mbox{$K^{-}  \pi^{+}$} system. 
  The longitudinal and parallel polarisation fractions are determined to be 
  \mbox{$f_{0} = 0.534 \pm 0.012 \pm 0.009$} 
  and 
  \mbox{$f_{\parallel} = 0.211 \pm 0.014 \pm 0.005$},
  respectively,
  where the first uncertainty is statistical and the second is systematic.
  The $C\!P$ asymmetries are measured with $3$--$7\%$ precision and are found to be consistent with zero. 
  These measurements, along with an updated determination of the branching fraction relative to the \mbox{$B^0  \rightarrow  J\mskip -3mu/\mskip -2mu\psi  K^{*0}$} decay, 
  are combined with previous \mbox{LHCb} results, providing the most precise values for these observables to date.
\end{abstract}

% % Plain Latex
% \begin{abstract}
% \noindent
%    A time-integrated angular analysis of the decay $B^0_s  \rightarrow  J/\psi  \overline{K}{}^{*}\kern-1pt(892)^{0}$, with $J/\psi  \rightarrow  \mu^{+} \mu^{-}$ and $\overline{K}{}^{*}\kern-1pt(892)^{0}  \rightarrow  K^{-}  \pi^{+}$, is presented. The analysis employs a sample of proton-proton collision data collected by the LHCb experiment during 2015-2018 at a centre-of-mass energy of $13 \text{TeV}$, corresponding to an integrated luminosity of $6 \text{fb}^{-1}$. A simultaneous maximum-likelihood fit is performed to the angular distributions in bins of the $K^{-} \pi^{+}$ mass. This fit yields measurements of the $CP$-averaged polarisation fractions and $CP$ asymmetries for the P-wave component of the $K^{-} \pi^{+}$ system. The longitudinal and parallel polarisation fractions are determined to be $f_{0} = 0.534 \pm 0.012 \pm 0.009$ and $f_{\parallel} = 0.211 \pm 0.014 \pm 0.005$, respectively, where the first uncertainty is statistical and the second is systematic. The $CP$ asymmetries are measured with $3$-$7\%$ precision and are found to be consistent with zero. These measurements, along with an updated determination of the branching fraction relative to the $B^0  \rightarrow  J/\psi  K^{*0}$ decay, are combined with previous LHCb results, providing the most precise values for these observables to date.
% \end{abstract}

% % Plain Latex
% \def\papertitle{
%     Updated measurement of $CP$ violation and polarisation in $B^0_s  \rightarrow  J/\psi \overline{K}{}^{*}\kern-1pt(892)^{0}$ decays
% }

\vspace*{1.5cm}

\begin{center}
  % To be submitted to
  % Submitted to
  Published in 
  JHEP 10 (2025) 173
  % JHEP
  % Phys.~Rev.~D /
  % Phys.~Rev.~Lett. /
 % Phys.~Lett.~B
  % Eur.~Phys.~J.~C /
  %  Nucl.~Phys.~B /
  % Chin.~Phys.~C /
  % Nature~Physics /
  % sciPost~Physics /
  % J. Instr. /
  % Instruments 
\end{center}

\vspace{\fill}

{\footnotesize 
% Edit macro in main.tex to keep metadata correct
\centerline{\copyright~\papercopyright. \href{\paperlicenceurl}{\paperlicence}.}}
\vspace*{2mm}

\end{titlepage}

%%%%%%%%%%%%%%%%%%%%%%%%%%%%%%%%
%%%%%  EOD OF TITLE PAGE  %%%%%%
%%%%%%%%%%%%%%%%%%%%%%%%%%%%%%%%

%  empty page follows the title page ----
\newpage
\setcounter{page}{2}
\mbox{~}
%\newpage
%
%% Author List ----------------------------
%%  You need to get a new author list!
%\input{LHCb_authorlist.tex}
%
%The author list for journal publications is provided by the Membership Committee shortly after 'approval to go to paper' has been given.
%%It will be made available on the page
%%\verb!http://www.physik.uzh.ch/~strauman/forMemCo/LHCb-PAPER-XXXX-XXX/! .
%It will be sent to you by email shortly after a paper number has beens assigned.
%The author list should be included already at first circulation, 
%to allow new members of the collaboration to verify whether they have been included correctly.
%Occasionally a misspelled name is corrected or associated institutions become full members.
%In that case, a new author list will be sent to you.
%In case line numbering doesn't work well after including the authorlist, try moving the \verb!\bigskip! after the last author to a separate line.
%
%
%The authorship for Conference Reports should be ``The LHCb
%  collaboration'', with a footnote giving the name(s) of the contact
%  author(s), but without the full list of collaboration names.

%\twocolumn
% %%%%%%%%%%%%% ---------

\renewcommand{\thefootnote}{\arabic{footnote}}
\setcounter{footnote}{0}

%%%%%%%%%%%%%%%%%%%%%%%%%%%%%%%%
%%%%%  Table of Content   %%%%%%
%%%%%%%%%%%%%%%%%%%%%%%%%%%%%%%%
%%%% Uncomment if desired
%\tableofcontents

\cleardoublepage

%%%%%%%%%%%%%%%%%%%%%%%%%
%%%%% Main text %%%%%%%%%
%%%%%%%%%%%%%%%%%%%%%%%%%

\pagestyle{plain} % restore page numbers for the main text
\setcounter{page}{1}
\pagenumbering{arabic}

%% Uncomment during review phase. 
%% Comment before a final submission.
% \linenumbers

%% This is the main body
%% It is useful to have a single file so comments are not missed in overleaf.
\section{Introduction}
\label{sec:Introduction}

The study of \CP violation in the \Bs meson system provides crucial tests of the Cabibbo--Kobayashi--Maskawa (CKM) mechanism within the Standard Model (SM) and enables investigation of potential contributions from physics beyond the Standard Model (BSM)~\cite{LHCb-PAPER-2011-021,DeBruyn:2022zhw}. A key observable is the \CP-violating phase \phis, which arises from the interference between \Bs meson decays proceeding directly to a \CP eigenstate and those occurring after \Bs--\Bsb mixing. Within the SM, neglecting subleading contributions, this phase is predicted to satisfy \mbox{$\phis \approx -2\beta_s$}, where \mbox{$\beta_s = \arg(-V_{ts}V_{tb}^* / V_{cs}V_{cb}^*)$} is related to elements of the CKM matrix~\cite{BIGI198185, PhysRevD.23.1567, PhysRevLett.45.952}. Global fits to experimental data, assuming CKM unitarity, provide precise indirect determinations of this phase~\cite{CKMfitter2015, UTfit-UT}.

Experimentally, \phis is measured directly using decays mediated by the \mbox{$\bquark \to \ccbar\squark$} transition, with the decay \mbox{\BsToJPsiPhi} serving as the benchmark channel. The current world average for the phase measured in \mbox{$\bquark \to \ccbar\squark$} transitions is \mbox{$\phi_s^{\ccbar\squark} = -0.052 \pm 0.013\rad$}~\cite{HFLAV23}. The excellent theoretical precision of the SM prediction and the relatively clean experimental signature make \phis a powerful probe for BSM effects, which could modify the \Bs--\Bsb mixing amplitude and shift the measured value relative to the SM expectation~\cite{Chiang2009NewPI}.

However, subleading SM contributions from electroweak loop (penguin) topologies in the \mbox{\BsToJPsiPhi} decay process can also induce a phase shift, $\Delta\phis$, which complicates the interpretation of the experimental measurement in terms of $\phis$ and thus BSM. The nonperturbative QCD nature of the penguin contributions makes theoretical calculations challenging~\cite{Liu:2013nea, Frings:2015eva}, which can be circumvented using data-driven strategies. 
One prominent approach, proposed in Ref.~\cite{Fleischer:1999zi} and further detailed in Refs.~\cite{Faller:2008gt, DeBruyn:2014oga, Barel:2020jvf, DeBruyn:2025rhk}, uses \CP asymmetry measurements of CKM-suppressed \mbox{$\bquark \to \ccbar\dquark$} decays, such as \mbox{\BsJpsiKst} and \mbox{\BdJpsiRho}, to constrain penguin effects in \mbox{\BsToJPsiPhi} via approximate \grpsuthree flavour symmetry. 
The LHCb collaboration has previously conducted such measurements for \mbox{\BsJpsiKstI}~\cite{LHCb-PAPER-2015-034} and \mbox{\BdJpsiRhoI}\cite{LHCb-PAPER-2014-058}, using
a data sample corresponding to an integrated luminosity of 
$3\invfb$, collected during 2011--2012 (referred to as Run~1). 
In addition, the LHCb collaboration also performed an \grpsuthree-symmetry-based analysis yielding constraints on $\Delta\phis$ with uncertainties ranging from 10 to 100~mrad depending on the inputs used~\cite{LHCb-PAPER-2015-034}. 
Improving the precision of the \mbox{\BsJpsiKst} parameters is crucial for refining these constraints, especially as the experimental uncertainty on $\phi_s^{\ccbar\squark}$ is reduced.

This paper presents an updated angular analysis of the decay \mbox{\Bs $\to$ (\jpsi $\to$ \mup\mun)(\Kstarzb $\to$ \Kmpip)} using the full dataset collected by the \lhcb experiment during the second LHC operational period (Run~2).\footnote{The inclusion of charge-conjugate processes is implied throughout, unless otherwise noted.} This dataset corresponds to an integrated luminosity of \mbox{$6\invfb$} of proton-proton (\proton\proton) collisions recorded at a centre-of-mass energy of \mbox{$13\tev$}.
The analysis focuses on the region dominated by the \KstarIzb resonance,
which is denoted as \Kstarzb for simplicity hereafter.
The decay rate is analysed as a function of three angles in the helicity basis (\costk, \costmu, \phihel),
which describe the orientation of the decay products in their respective rest frames.
This analysis follows the approach detailed in Ref.~\cite{LHCb-PAPER-2015-034} 
to determine the polarisation components, associated phase differences, and direct \CP asymmetries for both the \pwave (\Kstarzb vector meson) and \swave (spin-zero) components of the \Kmpip system.
The branching fraction relative to the \mbox{\Bd $\to$ \jpsi\Kstarz} decay is also measured, providing updated inputs to constrain penguin contributions in \mbox{$\bquark \to \ccbar\squark$} transitions.
Compared to the previous \lhcb result~\cite{LHCb-PAPER-2015-034}, the precision of this analysis can achieve benefits not only from a nearly twofold increase in the integrated luminosity, but also from an increased \bquark\bquarkbar production cross-section at the higher centre-of-mass energy, and from an optimised event selection that improves the sample purity. 
The larger data and simulation samples also allow for a significant reduction of the dominant systematic uncertainties, many of which were statistically limited in the previous measurement.

The remainder of this paper is organised as follows: a brief description of the \lhcb detector and the simulation samples used is provided in Sec.~\ref{sec:LHCbDetectorAndSimulation}, 
the reconstruction of candidate decays and the event selection criteria are detailed in Sec.~\ref{sec:ReconstructionAndSelection}, followed by the procedure for the fit to the mass distribution in Sec.~\ref{sec:FitToTheMassDistribution}. 
The methodology for the angular analysis is summarised in Sec.~\ref{sec:AngularAnalysis}, and the determination of the branching fraction is explained in Sec.~\ref{sec:BranchingRatioMeasurement}. 
The results obtained from the Run~2 data, along with their associated uncertainties, are presented in Sec.~\ref{sec:ResultsAndUncertainties}. 
These are then combined with previous \lhcb measurements from Run~1 in Sec.~\ref{sec:CombinationWithLHCbRun1Results}. 
Finally, Sec.~\ref{sec:Conslusions} provides the conclusions of this work.

% \FloatBarrier \newpage
\section{The \lhcb detector and simulation}
\label{sec:LHCbDetectorAndSimulation}

The \lhcb detector~\cite{LHCb-DP-2008-001, LHCb-DP-2014-002} is a single-arm forward
spectrometer covering the pseudorapidity range \mbox{$2 < \eta < 5$},
designed for the study of particles containing \bquark or \cquark quarks.
The detector includes a high-precision tracking system consisting of a silicon-strip vertex detector surrounding the \proton\proton interaction region~\cite{LHCb-DP-2014-001}, a large-area silicon-strip detector located upstream of a dipole magnet with a bending power of about \mbox{$4{\mathrm{\,T\,m}}$}, and three stations of silicon-strip detectors and straw drift tubes~\cite{LHCb-DP-2017-001} placed downstream of the magnet.
The tracking system provides a measurement of the momentum, \ptot, of charged particles with a relative uncertainty that varies from 0.5\% at low momentum to 1.0\% at \mbox{$200\gevc$}.
The minimum distance of a track to a primary \proton\proton collision vertex (PV), the impact parameter (IP), is measured with a resolution of \mbox{$(15 + 29/\pt)\mum$}, where \pt is the component of the momentum transverse to the beam, in \gevc.
Different types of charged hadrons are distinguished using information from two ring-imaging Cherenkov detectors (RICH)~\cite{LHCb-DP-2012-003}.
Photons, electrons and hadrons are identified by a calorimeter system consisting of scintillating-pad and preshower detectors, an electromagnetic and hadronic calorimeter. Muons are identified by a system composed of alternating layers of iron and multiwire proportional chambers~\cite{LHCb-DP-2012-002}.

Simulation samples are generated to model the signal (\BsJpsiKst) and normalisation (\BdJpsiKst) channels, as well as potential background contributions. Proton-proton collisions are generated using \pythia~8~\cite{Sjostrand:2007gs, *Sjostrand:2006za} with a specific \lhcb configuration~\cite{LHCb-PROC-2010-056}. Decays of hadronic particles are described by \evtgen~\cite{Lange:2001uf}, in which final-state radiation is generated using \photos~\cite{davidson2015photos}. The interaction of the generated particles with the detector and its response are implemented using the \geant toolkit~\cite{Allison:2006ve, *Agostinelli:2002hh} as described in Ref.~\cite{LHCb-PROC-2011-006}. In the signal and normalisation channel simulations, the \pwave components were generated to match the angular distributions observed in the \lhcb Run~1 analysis~\cite{LHCb-PAPER-2015-034}.

Several corrections to the simulated samples are applied to improve the agreement with data.
Differences in the \Bs/\Bd production kinematics and detector occupancy are corrected using a multivariate boosted decision tree (BDT) weighting technique~\cite{Rogozhnikov:2016bdp}. These corrections are derived from the high-statistics \mbox{\Bd $\to$ \jpsi\Kstarz} control channel using simulation and background-subtracted data, and then applied to both \Bd and \Bs simulated samples. 
Corrections for particle identification (PID) efficiencies are applied using calibration samples and a kernel density estimation method, accounting for track kinematics and detector occupancy~\cite{LHCb-PUB-2016-021, LHCb-DP-2018-001}. Further corrections related to angular acceptance are discussed in Sec.~\ref{subsec:AngularAcceptance}.

%\FloatBarrier
%\newpage
\section{Reconstruction and selection}
\label{sec:ReconstructionAndSelection}

Candidate \mbox{\Bs $\to$ \jpsi\Kstarzb} decays are reconstructed using \proton\proton collision data collected by the \lhcb detector during Run~2. The online event selection employs a trigger system~\cite{LHCb-DP-2012-004} consisting of a hardware stage followed by two software stages. The hardware trigger selects events based on high-\pt muons or high transverse energy calorimeter deposits; events passing any hardware decision are retained for this analysis. The first stage of the software trigger performs partial event reconstruction, selecting candidates primarily through signatures requiring displaced single muons, dimuons, or two-track vertices identified by multivariate algorithms~\cite{LHCb-PAPER-2019-042}. 
The second software stage performs a full event reconstruction, 
selecting \jpsi $\to$ \mup\mun candidates with good vertex-fit quality, significant displacement from any PV, and a dimuon mass within \mbox{$\pm 120 \mevcc$} of the known \jpsi mass~\cite{PDG2024}.

The offline selection criteria are optimised for the Run~2 dataset, following a strategy similar to that used in the Run~1 analysis~\cite{LHCb-PAPER-2015-034}.
The \jpsi candidates are formed from two oppositely charged tracks identified as muons using information from the muon chambers and RICH detectors. The two tracks must have a small distance of closest approach (DOCA) between their trajectories
and form a vertex with good vertex-fit quality. The dimuon mass is required to be within the range \mbox{$[-48.0, +42.9 ]\mevcc$} around the known \jpsi mass~\cite{PDG2024}.
The \Kstarzb candidates are formed from two oppositely charged tracks identified as a kaon and a pion using RICH information.
Similar requirements as for the muons are applied to DOCA and vertex-fit quality. 
The mass of the \mbox{\Kmpip} pair must be within \mbox{$\pm 70\mevcc$} of the known \Kstarzb mass~\cite{PDG2024}.
The \Bs candidates are formed by combining a \jpsi and a \Kstarzb candidate, requiring them to originate from a common vertex with a good vertex-fit quality. 

All four final-state tracks must be inconsistent with originating from any PV. 
The \Bs decay vertex must be significantly displaced from its associated PV, defined as that which gives the smallest \chisqip for the \Bs candidate. Here, \chisqip is defined as the difference in the vertex-fit $\chi^2$ of a given PV reconstructed with and without the particle under consideration.
The cosine of the angle between the \Bs momentum vector and the vector pointing from the associated PV to the \Bs decay vertex (direction angle, DIRA) is required to be greater than 0.999.
To suppress candidates containing duplicate or partial track reconstructions, the minimum opening angle between any pair of final-state tracks is required to be greater than $0.5\mrad$.

A kinematic fit~\cite{Hulsbergen:2005pu} is performed to improve the resolution of the decay kinematics and the helicity angles. The fit constrains the dimuon mass to the known \jpsi mass and requires the \Bs candidate to originate from its associated PV.

\subsection{Combinatorial background}
\label{sec:CombinatorialBackground}

A BDT classifier~\cite{Breiman} with gradient boosting is used to suppress the large combinatorial background remaining after the initial selection. 
To account for variations in data-taking conditions, separate BDTs are trained for three distinct periods: 2015--2016, 2017, and 2018. Simulated \mbox{\Bs $\to$ \jpsi\Kstarzb} events, corrected as described in Sec.~\ref{sec:LHCbDetectorAndSimulation}, serve as the signal proxy. Data candidates from the high-mass sideband, \mbox{$m(\jpsi\Kmpip) \in [5440, 5650]\mevcc$}, serve as the background proxy. The $k\text{-folding}$ technique (\mbox{$k=10$})~\cite{kFold_modified} is used during training. 
Fourteen variables, selected for their ability to discriminate between signal and combinatorial background, are used as inputs to the BDT. These include:
\begin{itemize}
    \item Track variables: minimum track \chisqip with respect to any PV, maximum DOCA between final-state tracks, maximum track fit \chisqndf (chi-squared per degree of freedom) for hadron and muon tracks separately;
    \item Vertex variables: vertex-fit \chisqndf of the \jpsi and \Bs decay vertices;
    \item Particle identification: variables quantifying the confidence of muon identification, and neural network probabilities indicating whether hadron tracks correspond to kaons or pions;
    \item Kinematics: transverse momentum of the \Bs candidate;
    \item Flight information: \Bs candidate proper decay time, \Bs \chisqip relative to its associated PV, \Bs DIRA;
    \item Fit quality: \chisq from the \Bs kinematic fit.
\end{itemize}
The trained classifiers are applied to the respective data and simulation samples, 
and a selection requirement is applied to the BDT output for each data-taking period. 
This requirement is chosen to maximise the statistical precision of the angular analysis (Sec.~\ref{sec:AngularAnalysis}), which involves balancing the retention of signal events against the rejection of background.
% while maintaining similar signal efficiency across the years.
This selection retains approximately 92.7\% of the signal candidates passing the preceding requirements, while rejecting around 95.6\% of the combinatorial background.
Multiple candidates from the same \proton\proton bunch crossing are rare (\mbox{$\sim 0.05\%$}) after this selection; all such candidates are retained~\cite{Koppenburg:2017zsh}.

\subsection{Peaking background}
\label{sec:PeakingBackground}

Specific background sources involving misidentified particles or partially reconstructed decays can form peaking structures in the \mbox{$m(\jpsi\Kmpip)$} spectrum, potentially biasing both the signal yield and angular distributions. These contributions are evaluated and suppressed.
The backgrounds considered include \mbox{\Bu $\to$ \jpsi\Kp}, \mbox{\Bd $\to$ \jpsi\pip\pim}, \mbox{\Bs $\to$ \jpsi\pip\pim}, \mbox{\Bs $\to$ \jpsi\Kp\Km}, \mbox{\Lb $\to$ \jpsi\proton\Km}, and \mbox{\Lb $\to$ \jpsi\proton\pim} decays.
These backgrounds are studied using simulated samples, and their expected yields are evaluated taking into account the sample luminosity, selection efficiencies, production cross-section from Ref.~\cite{LHCb-PAPER-2016-031}, fragmentation fractions from Ref.~\cite{LHCb-PAPER-2020-046}, and decay branching fractions from Ref.~\cite{PDG2024}.

Dedicated vetoes based on alternative mass hypotheses are applied to suppress the dominant background contributions.
These include \mbox{\Bu $\to$ \jpsi\Kp} decays where the \jpsi and \Kp are combined with an unrelated pion, and decays such as \mbox{\LbJpsipK} and \mbox{\LbJpsippi} involving particle misidentification. 
A combination of mass and PID requirements is used to reduce contributions from \mbox{\BdJpsipipi} and \mbox{\BsJpsipipi} decays. 

After these steps, following the Run~1 analysis approach~\cite{LHCb-PAPER-2015-034}, the remaining small contributions (at the level of 1\% of the signal yield) from \mbox{\BdJpsipipi}~\cite{LHCb-PAPER-2014-012}, \mbox{\BsJpsiKK}~\cite{LHCb-PAPER-2012-040}, and \mbox{\LbJpsipK}~\cite{LHCb-PAPER-2015-029} decays are statistically subtracted. 
This is achieved by adding simulated decays for these channels to the data sample, using negative weights. These simulated events are added prior to the mass fit (Sec.~\ref{sec:FitToTheMassDistribution}), weighted according to their known amplitude models and scaled to their estimated residual yields.
The remaining background contributions are sufficiently small and therefore neglected.

\subsection{Candidates with degraded momentum}
\label{sec:RemovalOfPoorlyReconstructedCandidates}

Decays in which final-state hadrons decay in flight or interact with detector material can lead to poorly measured momenta and contribute to tails in the reconstructed \mbox{$m(\jpsi\Kmpip)$} distribution. This effect is particularly relevant for the \Bd control channel, where the tail extends significantly into the \Bs signal region. Approximately 3\% of reconstructed \Bd candidates have masses above \mbox{$5325\mevcc$}, contaminating the \Bs peak.
To suppress these contributions, a dedicated BDT classifier is trained using simulated \mbox{\Bd $\to$ \jpsi\Kstarz} events. Signal proxy events are those where both kaon and pion tracks reach at least the second tracking station downstream of the magnet without significant interaction or decay. Background proxy events are those where at least one hadron fails this condition. Nine input variables are used related to track-fit quality, consistency between tracking subsystems, ghost probability~\cite{DeCian:2255039}, the probability of being misidentified as muons, and track momenta. After applying simulation corrections and $k\text{-folding}$ (\mbox{$k=10$}), a loose requirement on the BDT output is applied to data and simulation. 
This selection significantly reduces the tail from poorly reconstructed \Bd decays. 
Specifically, the ratio of \Bd candidates falling within the \Bs mass window (defined as \mbox{$\pm 30\mevcc$} around the known \Bs mass~\cite{PDG2024}) to the number of \Bs signal candidates in the same window is decreased from approximately 12.8\% to 7.6\%. 
This reduction is achieved while maintaining a high signal efficiency of approximately 97.8\% for the \Bs decays.

%\FloatBarrier
%\newpage
\section{Fit to the mass distribution}
\label{sec:FitToTheMassDistribution}

The sample of selected \mbox{\jpsi\Kmpip} candidates, after the application of selection requirements (Sec.~\ref{sec:ReconstructionAndSelection}) and the statistical subtraction of known peaking backgrounds (Sec.~\ref{sec:PeakingBackground}), consists primarily of signal \mbox{\Bs $\to$ \jpsi\Kstarzb} decays, control channel \mbox{\Bd $\to$ \jpsi\Kstarz} decays, and combinatorial background. 
To determine the yields of the signal and control channels, as well as combinatorial background, an extended unbinned maximum-likelihood fit is performed to the \mbox{$m(\jpsi\Kmpip)$} mass distribution. 
From the results of this fit, \sPlot weights~\cite{Pivk:2004ty} are calculated, using \mbox{$m(\jpsi\Kmpip)$} as the discriminating variable. These weights are then used to statistically separate the different signal and background components, providing the necessary background-subtracted distributions for the subsequent angular analysis (Sec.~\ref{sec:AngularAnalysis}).

To account for variations across data-taking periods and the dependence of the signal shape on the \mbox{\Kmpip} mass (\mkpi), the data sample is divided into twelve subsamples. These are based on three data-taking periods (2015--2016, 2017, 2018) and four bins in \mkpi, each \mbox{$35\mevcc$} wide and spanning the region centred around the known \Kstarzb mass~\cite{PDG2024}. 
The fit is performed simultaneously across these subsamples.

The probability density function (PDF) used in the fit describes three main components: \Bd and \Bs signals, and combinatorial background.
The \Bd and \Bs signal shapes are modelled using a customised double-sided Crystal Ball function (2DSCB). 
This function offers a description of the signal tails comparable to that of the Hypatia function used in the Run~1 analysis~\cite{LHCb-PAPER-2015-034}, particularly accounting for effects such as final-state radiation from muons (bremsstrahlung), while providing improved robustness in the fit minimisation process.
The 2DSCB function is constructed as a sum of two DSCB components~\cite{Skwarnicki:1986xj}, 
sharing the same peak position $\mu$ and tail parameters ($\alpha_L, \alpha_R, n_L, n_R$), but with different resolution parameters ($\sigma_1, \sigma_2$) combined with a relative fraction $f_{1}$,
\begin{equation}
    \begin{aligned}
    \mathcal{P}_{\text{2DSCB}}(m; \vec{\vartheta}) = & f_{1} \mathcal{P}_{\text{DSCB}}(m; \alpha_{L}, \alpha_{R}, n_{L}, n_{R}, \sigma_{1}, \mu) \\
    & + (1 - f_{1}) \mathcal{P}_{\text{DSCB}}(m; \alpha_{L}, \alpha_{R}, n_{L}, n_{R}, \sigma_{2}, \mu),
    \end{aligned}
    \label{eq:doubleDSCB}
\end{equation}
where $\vec{\vartheta}$ represents the set of shape parameters. For both \Bd and \Bs signals, the tail parameters and the fraction $f_{1}$ are shared within each \mkpi bin and are fixed to values determined from fits to simulation. 
The position of the \Bs peak, $\mu_{\Bs}$, is constrained relative to the \Bd peak ($\mu_{\Bd}$) using the world-average mass difference $\Delta m$~\cite{PDG2024}, \ie \mbox{$\mu_{\Bs} = \mu_{\Bd} + \Delta m$}.
The core Gaussian resolutions for the 2DSCB function are determined from fits to simulated \Bd events and shared between \Bd and \Bs decays. 
To account for potential differences between data and simulation, these simulation-based resolutions are adjusted in the fit to data using multiplicative scale factors, $s_{\Bs}$ and $s_{\Bd}$, applied independently to the signal and normalisation channels, respectively. 
These scale factors ($s_{\Bs}$ and $s_{\Bd}$) and the \Bd peak position ($\mu_{\Bd}$) are allowed to vary freely in the fit, determined independently for each \mkpi bin but shared across the data-taking years.
Finally, the combinatorial background component is described by an exponential function, \mbox{$\mathcal{P}_{\text{comb}}(m) \propto e^{k_{\text{comb}} m}$}, where the slope parameter $k_{\text{comb}}$ is allowed to vary freely in the fit for each subsample.

The full PDF per subsample is given by
\begin{equation}
\begin{aligned}
    \mathcal{P}_{\text{full}}(m; \vec{\vartheta}, \mu_{\Bd}, \Delta m, s_{\Bd}, s_{\Bs})
        & = N_{\Bs}  \mathcal{P}_{\text{2DSCB}}(m; \vec{\vartheta}, \mu_{\Bd}+\Delta m, s_{\Bs}) \\
        & + N_{\Bd}  \mathcal{P}_{\text{2DSCB}}(m; \vec{\vartheta}, \mu_{\Bd}, s_{\Bd}) \\
        & + N_{\text{comb}} \mathcal{P}_{\text{comb}}(m; k_{\text{comb}}),
\end{aligned}
\label{eq:fullpdf}
\end{equation}
where $\vec{\vartheta}$ denotes the shape parameters that are fixed to values determined from simulation, 
while the yields $N_{\Bs}$, $N_{\Bd}$, and $N_{\text{comb}}$ are free parameters in each subsample.

\begin{figure}[htbp]
    \centering
    \includegraphics[width=0.49\textwidth] {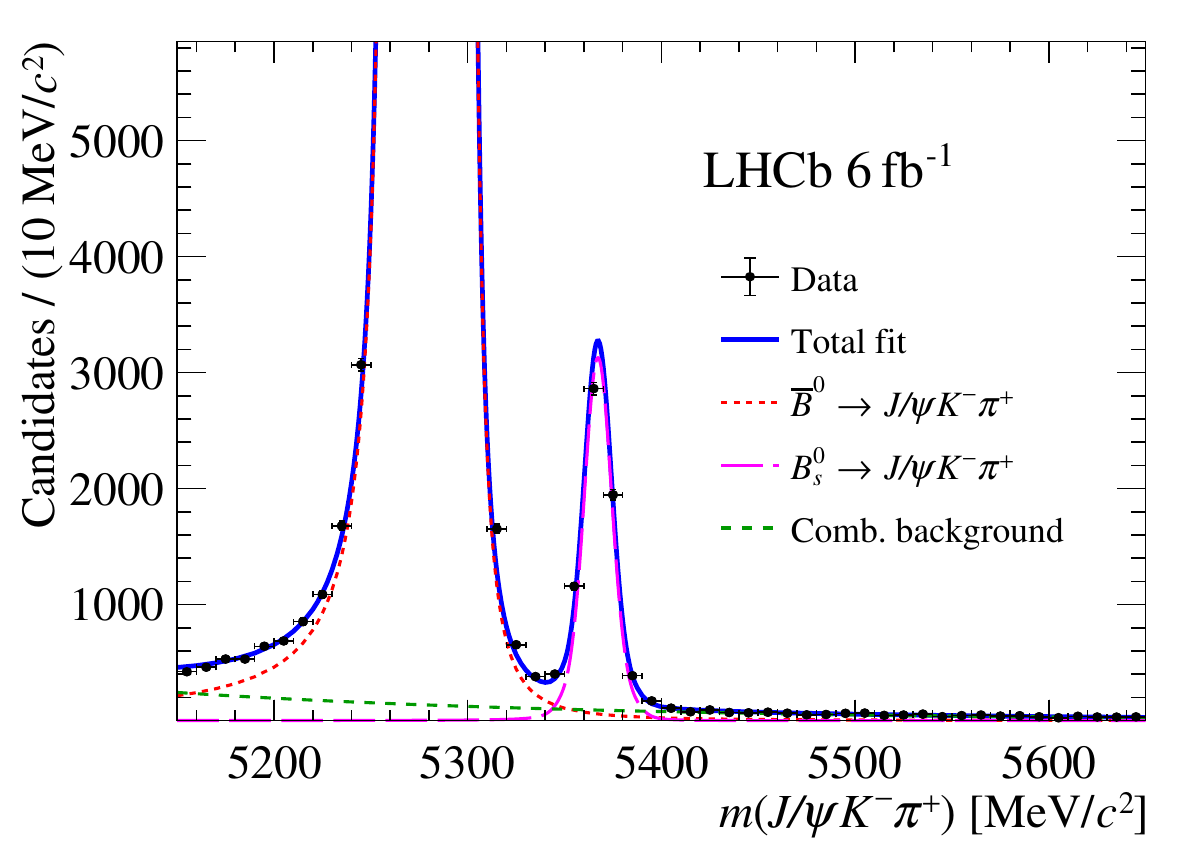}
    \includegraphics[width=0.49\textwidth] {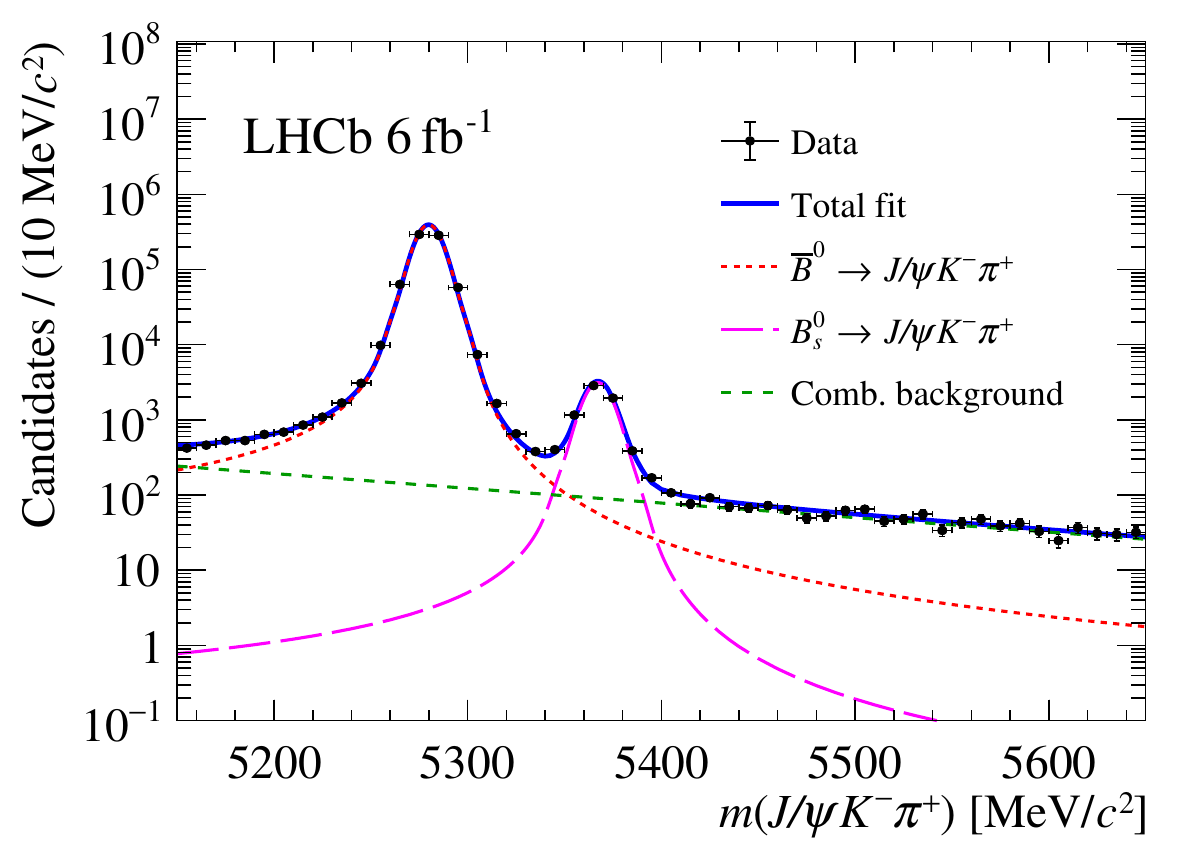}
    \caption{
        Projection of the simultaneous fit to the 
        \mbox{$m(J\mskip -3mu/\mskip -2mu\psi  K^{-}  \pi^{+})$} 
        distribution for the full Run~2 data sample, summed over data-taking periods and $m_{K\pi}$ bins. 
        The distribution is shown with a (left) linear y-axis scale, where the vertical range is adjusted to highlight the $B^0_s$ signal, 
        and a (right) logarithmic y-axis scale to display the full range. 
       The data points and the total fit function, along with its components, are described in the legend.
   }
    \label{fig:massfit:final:Data:Projection:allYears}
\end{figure}

The simultaneous fit is performed across all twelve subsamples in the mass range \mbox{$5150 < m(\jpsi\Kmpip) < 5650\mevcc$}. Projections of the fit result, summed over all subsamples, are shown in Fig.~\ref{fig:massfit:final:Data:Projection:allYears}.
The overall yields, summed across all \mkpi bins and years, are determined to be
\begin{align*}
    N_{\Bd} &= 726\,540 \pm 860 \stat \pm 260 \syst, \\
    N_{\Bs} &= \phantom{00}6\,098 \pm \phantom{0}84 \stat \pm \phantom{0}39 \syst,
\end{align*}
where the systematic uncertainties arise primarily from the choice of fit model (Sec.~\ref{sec:ResultsAndUncertainties}). 
The ratio of these fitted yields, which is a key ingredient in the branching fraction measurement (Sec.~\ref{sec:BranchingRatioMeasurement}), is found to be
\begin{equation*}
    \frac{N_{\Bs}}{N_{\Bd}} = ( 8.39 \pm 0.12\stat \pm 0.05\syst)\times 10^{-3},
\end{equation*}
where the small correlations between \Bd and \Bs yields are taken into account.

% --- Figure Placement ---
% --- End Figure Placement ---

Based on the fit results within each subsample, \sPlot weights~\cite{Pivk:2004ty} are calculated to statistically separate the signal, normalisation, and background components for the angular analysis. 
A key assumption for the unbiased application of the \sPlot technique is the lack of significant correlation between the discriminating variable (\mbox{$m(\jpsi\Kmpip)$}) and the variables of interest for the subsequent analysis (the angles: \costk, \costmu, \phihel). 
This lack of correlation has been verified using the Kendall rank correlation test~\cite{Forthofer1981, COWs} on simulation and data sidebands (\textit{p}-values > 0.1).
Residual correlations are accounted for using input from simulation, and considered when evaluating systematic uncertainties (Sec.~\ref{sec:ResultsAndUncertainties}).

%\FloatBarrier
%\newpage
\section{Angular analysis}
\label{sec:AngularAnalysis}

The polarisation fractions, phase differences, and direct \CP asymmetries of the \mbox{\Bs $\to$ \jpsi\Kstarzb} decay are extracted via an unbinned maximum-likelihood fit to the three helicity angles \mbox{$\Omega_{\text{hel}} = (\costk, \costmu, \phihel)$}. The fit is performed on the background-subtracted data sample, obtained using the \sPlot weights derived from the mass fit~\cite{Xie:2009rka}. 
The analysis closely follows the methodology detailed in the previous \lhcb measurement using Run~1 data~\cite{LHCb-PAPER-2015-034}.

%\FloatBarrier 
\subsection{Angular formalism}
\label{subsec:AngularFormalism}

The decay is described in the helicity basis, where the helicity angles are defined as in Ref.~\cite{LHCb-PAPER-2015-034}. The analysis considers contributions from the spin-one (\pwave) \Kstarzb resonance and an effective spin-zero (\swave) \mbox{$K\pi$} component. The time-integrated decay rate can be expressed in the helicity basis as~\cite{Zhang2012TimedependentDF}
\begin{equation}
    \frac{\deriv \Gamma(\Omega_{\text{hel}})}{\deriv \Omega_{\text{hel}}} 
    \propto \sum_{\alpha_{\mu} = \pm1} \Bigg|
    \sum_{\lambda,J}^{|\lambda|\leq J}\sqrt{\frac{2J+1}{4\pi}}\mathcal{H}_{\lambda}^{J}
    e^{-i\lambda\phihel}d_{\lambda,\alpha_{\mu}}^{1}(\thetamu)d_{-\lambda,0}^{J}(\thetaK) \Bigg|^2 ,
    \label{eq:helPDF}
\end{equation}
where \mbox{$\lambda = 0, \pm1$} is the \jpsi helicity, \mbox{$\alpha_{\mu} = \pm1$} is the difference between the helicities of the two muons, $J$ is the spin of the \mbox{$K\pi$} system (\mbox{$J=0$} for \swave, \mbox{$J=1$} for \pwave), $\mathcal{H}_{\lambda}^{J}$ are the helicity amplitudes, and $d^j_{m'm}$ are the Wigner small \textit{d}-matrices.

Working in the transversity basis, the amplitudes ($A_0, A_{\parallel}, A_{\perp}, A_S$) have definite \CP properties, related to the helicity amplitudes via
\begin{equation}
    \begin{aligned}
        A_{0} &= \mathcal{H}^1_0, \\
        A_{\parallel} &= \frac{1}{\sqrt{2}} (\mathcal{H}^1_{+1} + \mathcal{H}^1_{-1}), \\
        A_{\perp} &= \frac{1}{\sqrt{2}} (\mathcal{H}^1_{+1} - \mathcal{H}^1_{-1}), \\
        A_S &= \mathcal{H}^0_0.
    \end{aligned}
    \label{eq:AmpTransHel}
\end{equation}
These transversity amplitudes are complex numbers, $A_k = |A_k| e^{i\delta_k}$ (for $k=0, \parallel, \perp, S$), where $|A_k|$ is the magnitude and $\delta_k$ is the corresponding phase. For the \CP-conjugate decay \mbox{\Bsb $\to$ (\jpsi $\to$ \mup\mun)(\decay{\Kstarz}{ \Kp\pim})}, the corresponding amplitudes are denoted $\overline{A}_k$.

The differential decay rate can be expanded as a sum of ten terms involving products ($a_k$) of these amplitudes multiplied by specific angular functions $G_k(\Omega_{\text{hel}})$,
\begin{equation}
    \frac{\mathrm{d} \Gamma(\Omega_{\text{hel}})}{\mathrm{d} \Omega_{\text{hel}}}
        = \sum_{k=1}^{10} a_{k} G_k(\Omega_{\text{hel}}) .
    \label{eq:SumOfAngPDF}
\end{equation}
The terms $a_k$ involve bilinear products of the transversity amplitudes. Specifically, the terms with index $k=1, 2, 3, 7$ correspond to the squared magnitudes $|A_0|^2, |A_{\parallel}|^2, |A_{\perp}|^2, |A_S|^2$, respectively. The remaining terms ($k=4, 5, 6, 8, 9, 10$) represent interference between different transversity amplitudes. 
The explicit forms of $a_k$ and $G_k$ are given in Table~\ref{tab:angAna:decayDescription_SPwave}.
For the terms representing squared amplitudes, the table also shows the correspondence between the summation index $k$ and the respective polarisation state labels ($0, \parallel, \perp, S$).
This expansion is equivalent to that in Ref.~\cite{Zhang2012TimedependentDF}. 
A similar expansion exists for the \Bsb decay rate using the amplitudes $\overline{A}_k$.

% Definition of Physics Asymmetry ACP_k

The direct \CP asymmetry for a specific final-state polarisation $k$ is defined using the partial decay rates $\Gamma_k = \int a_k G_k(\Omega_{\text{hel}}) \mathrm{d}\Omega_{\text{hel}}$ integrated over angles (where $a_k$ involves only $|A_k|^2$ or $|\overline{A}_k|^2$ for $k=0,\parallel,\perp,S$),
\begin{equation}
    A^{\CP}_{k} = \frac{\overline{\Gamma}_k - \Gamma_k}{\overline{\Gamma}_k + \Gamma_k}
    \quad (k=0, \parallel, \perp, S).
    \label{eq:cpv:def:ACP_k_formalism_corrected}
\end{equation}

% Parameters extracted from the simultaneous fit
The analysis employs a simultaneous maximum-likelihood fit across the eight categories defined by the four \mkpi bins (Sec.~\ref{sec:FitToTheMassDistribution}) and the two meson states (\Bs and \Bsb), tagged by the reconstructed kaon charge. 
This fit determines the physics asymmetries $A^{\CP}_k$ defined in Eq.~\ref{eq:cpv:def:ACP_k_formalism_corrected}, along with parameters describing the \CP-averaged decay dynamics.
Introducing the \CP-averaged partial decay rate for polarisation state $k$ as $\Gamma_{k}^{\text{avg}} = (\Gamma_k + \overline{\Gamma}_k)/2$, the \CP-averaged parameters extracted by the fit are:
\begin{itemize}
    \item The \pwave polarisation fractions $f_k$ ($k=0, \parallel, \perp$).
    These represent the fraction of the total \CP-averaged \pwave decay rate ($\Gamma_{\text{P}}^{\text{avg}} = \Gamma_{0}^{\text{avg}} + \Gamma_{\parallel}^{\text{avg}} + \Gamma_{\perp}^{\text{avg}}$) associated with each polarisation state,
    \begin{equation} % Using equation* for unnumbered definition
      f_k = \frac{\Gamma_{k}^{\text{avg}}}{\Gamma_{\text{P}}^{\text{avg}}} = \frac{\Gamma_{k}^{\text{avg}}}{\Gamma_{0}^{\text{avg}} + \Gamma_{\parallel}^{\text{avg}} + \Gamma_{\perp}^{\text{avg}}}.
    \end{equation}
    By definition, these fractions satisfy the constraint $f_0 + f_\parallel + f_\perp = 1$.

    \item The \swave fraction $F_S$.
    This represents the fraction of the total \CP-averaged decay rate ($\Gamma^{\text{avg}}_{\text{tot}} = \Gamma_{\text{P}}^{\text{avg}} + \Gamma_{S}^{\text{avg}}$) originating from the \swave component,
    \begin{equation} % Using equation* for unnumbered definition
      F_S = \frac{\Gamma_{S}^{\text{avg}}}{\Gamma^{\text{avg}}_{\text{tot}}} = \frac{\Gamma_{S}^{\text{avg}}}{\Gamma_{0}^{\text{avg}} + \Gamma_{\parallel}^{\text{avg}} + \Gamma_{\perp}^{\text{avg}} + \Gamma_{S}^{\text{avg}}}.
    \end{equation}

    \item The phase differences $\delta_{\parallel} - \delta_0$, $\delta_{\perp} - \delta_0$, and $\delta_{S} - \delta_0$, 
    adopting the convention $\delta_0 = 0$.
\end{itemize}
The detailed parameterisation implemented in the likelihood function to extract these quantities is discussed further in Sec.~\ref{subsec:CPAsymmetries}.

% Fit configuration details
Within the simultaneous fit, the \pwave parameters ($f_0, f_\parallel, \delta_\parallel, \delta_\perp, A^{\CP}_0, A^{\CP}_\parallel, A^{\CP}_\perp$) and the \swave \ \CP asymmetry ($A^{\CP}_S$) are constrained to be the same across all \mkpi bins. The remaining \swave parameters, the fraction $F_S$ and the phase $\delta_S$, are allowed to vary independently in each bin. 
The mass dependence of the interference between the \swave and \pwave components is accounted for by scaling the interference terms ($k=8, 9, 10$ in Eq.~\ref{eq:SumOfAngPDF}) by the complex factor $C_{\rm{SP}} e^{-i\Theta_{\rm{SP}}}$, which depends on the \Kpi mass spectrum. The same numerical values for $C_{\rm{SP}}$ are used as in Ref.~\cite{LHCb-PAPER-2015-034}, and the interference phase $\Theta_{\rm{SP}}$ is absorbed into the definition of the fitted S-wave phase $\delta_S$.

% Table placement might need adjustment
\begin{table}[!bt]
    \caption{Angular basis functions $G_k(\Omega_{\text{hel}})$ and the corresponding amplitude products $a_k$ appearing in the differential decay rate Eq.~\ref{eq:SumOfAngPDF}, as defined in Ref.~\cite{Zhang2012TimedependentDF}.
    Also shown in parentheses in the first column are the corresponding transversity basis labels for the squared-amplitude terms.}
    \begin{center}
    \resizebox{\textwidth}{!}{%
        {\renewcommand{\arraystretch}{1.5} % Apply line spacing only to this table
            \begin{tabular}{lcc}
                \hline
                Index $k$ & Amplitude product $a_k$ & Angular function $G_k(\Omega_{\text{hel}})$ \\
                \hline
                $\phantom{0}1$ ($0$) & $|A_0|^2$ & $\frac{9}{16\pi} \cosSqtk (1 -\cosSqtmu)$ \\
                $\phantom{0}2$ ($\parallel$) & $|A_{\parallel}|^2$ & $\frac{9}{32\pi} (1 - \cosSqtk) (1 - (1 -\cosSqtmu) \cos^2\phihel)$ \\
                $\phantom{0}3$ ($\perp$) & $|A_{\perp}|^2$ & $\frac{9}{32\pi} (1 - \cosSqtk) ((1 -\cosSqtmu) \cos^2\phihel +\cosSqtmu)$ \\
                % \hline
                $\phantom{0}4$ & $\Imag(A_{\parallel}^* A_{\perp})$ & $\frac{9}{16\pi} (1 - \cosSqtk) (1 -\cosSqtmu) \sin(2\phihel)$ \\
                $\phantom{0}5$ & $\Real(A_0^* A_{\parallel})$ & $\frac{9\sqrt{2}}{16\pi} \costk \costmu \sqrt{(1 - \cosSqtk)(1 -\cosSqtmu)} \cos\phihel$ \\
                $\phantom{0}6$ & $\Imag(A_0^* A_{\perp})$ & $\frac{9\sqrt{2}}{16\pi} \costk \costmu \sqrt{(1 - \cosSqtk)(1 -\cosSqtmu)} \sin\phihel$ \\
                % \hline
                $\phantom{0}7$ ($S$) & $|A_S|^2$ & $\frac{3}{16\pi} (1 -\cosSqtmu)$ \\
                % \hline
                $\phantom{0}8$ & $\Real(A_S^* A_{\parallel})$ & $\frac{3\sqrt{6}}{16\pi} \costmu \sqrt{(1 - \cosSqtk)(1 -\cosSqtmu)} \cos\phihel$ \\
                $\phantom{0}9$ & $\Imag(A_S^* A_{\perp})$ & $\frac{3\sqrt{6}}{16\pi} \costmu \sqrt{(1 - \cosSqtk)(1 -\cosSqtmu)} \sin\phihel$ \\
                10 & $\Real(A_S^* A_0)$ & $\frac{3\sqrt{3}}{8\pi} \costk (1 -\cosSqtmu)$ \\
                \hline
            \end{tabular}
        } % resize
    } % \renewcommand
    \end{center}
    \label{tab:angAna:decayDescription_SPwave}
\end{table}

%\FloatBarrier 
\subsection{Angular acceptance}
\label{subsec:AngularAcceptance}

The detector geometry and selection criteria distort the observed angular distributions relative to the true decay rate described by Eq.~\ref{eq:SumOfAngPDF}.
This acceptance effect, represented by the function $\varepsilon(\Omega_{\text{hel}})$, must be accounted for in the fit. 
The acceptance function is determined using simulated samples.
To ensure these samples accurately reflect Run~2 data conditions, several corrections are applied to account for potential imperfections in simulating the detector response and particle kinematics.
These corrections involve both kinematic weighting and an iterative procedure designed to ensure that the angular distributions in the corrected simulation match those observed in data, following strategies similar to previous \lhcb analyses~\cite{LHCb-PAPER-2015-034, LHCb-PAPER-2019-013}. 
The kinematic variables corrected via weighting include the total and transverse momentum of the \Bs meson and the final-state kaon and pion tracks, and the $K\pi$ mass. 
In the baseline approach, corrections for the \Bs and final-state track kinematics are primarily derived using the \Bd control channel, 
while the \mkpi correction, which weights the simulated \mkpi distribution to match that from background-subtracted \Bs data, is derived using the \Bs signal sample.

The acceptance correction is incorporated into the likelihood fit using the method of normalisation weights~\cite{PhysRevLett.87.241801, BaBar:2004xhu}. This method involves calculating the integrals
\begin{equation}
    \xi_k = \int \varepsilon(\Omega_{\text{hel}}) G_k(\Omega_{\text{hel}}) \mathrm{d} \Omega_{\text{hel}}
    \label{eq:ACC:modelling:NW:def}
\end{equation}
for each angular function $G_k$ (defined in Table~\ref{tab:angAna:decayDescription_SPwave}) using the corrected simulation samples.
These weights $\xi_k$ are calculated separately for each of the eight analysis subsamples (4 \mkpi bins $\times$ 2 kaon charges) and are used directly in the construction of the likelihood function, obviating the need for an explicit parameterisation of $\varepsilon(\Omega_{\text{hel}})$. 
An example set of the normalisation weights is shown in Table~\ref{tab:angAccWeightsUncorr}. 
Statistical uncertainties arising from the finite size of the simulation samples used to calculate the $\xi_k$ weights are evaluated using bootstrap replicas, following standard resampling techniques~\cite{efron:1979}, and propagated as systematic uncertainties on the measured physics parameters.

% Table placement might need adjustment
\begin{table}[!bt]
  \caption{Example normalisation weights $\xi_k$ determined for Run~2 \Bs candidates in the third $m_{\Kpi}$ bin with a \Km meson, normalised such that $\xi_1=1$. The index $k$ corresponds to the terms in Table~\ref{tab:angAna:decayDescription_SPwave}. Uncertainties are statistical only, arising from the limited size of the simulation sample.}
  \center
  \begin{tabular}{ccc}
    \hline
    Index $k$ & Amplitude product $a_k$& Normalised weight $\xi_k / \xi_1$ \\
    \hline
     1  & $|A_0|^2$                                                               & $\phantom{-}1             $\\
     2  & $|A_{\parallel}|^2$                                                     & $\phantom{-}1.4249 \pm  0.0073$   \\
     3  & $|A_{\perp}|^2$                                                         & $\phantom{-}1.4489 \pm  0.0074$   \\
     4  & $|A_{\parallel}| |A_{\perp}| \sin(\delta_{\parallel} - \delta_{\perp})$ & $-0.0013 \pm  0.0043$   \\
     5  & $|A_0| |A_{\parallel}| \cos(\delta_{\parallel})$                        & $-0.0189 \pm  0.0029$   \\
     6  & $|A_0| |A_{\perp}| \sin(\delta_{\perp})$                                & $\phantom{-}0.0027 \pm  0.0028$   \\
     7  & $|A_S|^2$                                                               & $\phantom{-}1.2485 \pm  0.0062$   \\
     8  & $|A_S| |A_{\parallel}| \cos(\delta_S - \delta_{\parallel})$             & $-0.0412 \pm  0.0040$   \\
     9  & $|A_S| |A_{\perp}| \sin(\delta_S - \delta_{\perp})$                     & $\phantom{-}0.0072 \pm  0.0040$   \\
     10 & $|A_S| |A_0| \cos(\delta_S)$                                            & $ -0.6919 \pm  0.0096$   \\
    \hline
  \end{tabular}
  \label{tab:angAccWeightsUncorr}
\end{table}

%\FloatBarrier 
\subsection{\boldmath \CP asymmetries}
\label{subsec:CPAsymmetries}

% Connect to the maximum-likelihood fit
The maximum-likelihood fit described in Sec.~\ref{subsec:AngularFormalism} aims to determine the direct physics \CP asymmetries $A^{\CP}_k$ (defined in Eq.~\ref{eq:cpv:def:ACP_k_formalism_corrected}),
alongside the \CP-averaged polarisation fractions ($f_k, F_S$) and phases ($\delta_k$), which were introduced and defined in Sec.~\ref{subsec:AngularFormalism}. 
These are the fundamental physics parameters extracted from the data.

% Internal parameterization used in the fit model
To achieve this, the likelihood function models the decay rates for \Bs and \Bsb candidates separately, using a parameterisation based on these physics parameters. 
The \CP-averaged polarisation fractions $f_k$ ($k=0,\parallel,\perp$) and $F_S$, as defined by their relation to the \CP-averaged partial decay rates in Sec.~\ref{subsec:AngularFormalism}, correspond to the average contributions of each polarisation state.
These correspond to squared average amplitudes, 
denoted $|A_k|^2_{\text{avg}}$.
Specifically, $|A_k|^2_{\text{avg}}$ is equivalent to $f_k$ for P-waves, and for the \swave, $|A_S|^2_{\text{avg}}$ is related to $F_S$ via $F_S = |A_S|^2_{\text{avg}} / (1+|A_S|^2_{\text{avg}})$.

The overall normalisation yields for the physics decays \Bs (\mbox{$N(\Bs \to f)$}) and \Bsb (\mbox{$N(\Bsb \to \overline{f})$}) are parameterised in terms of an effective average total yield $N_{\text{avg}}$ (also determined by the fit) and an effective physics \CP asymmetry \mbox{$\mathbb{A}^{\CP}_{\text{phys}} = \sum_k |A_k|^2_{\text{avg}} A^{\CP}_k /  (1+|A_S|^2_{\text{avg}})$},
\begin{equation}
    N(\Bs \to f) = N_{\text{avg}} (1 - \mathbb{A}^{\CP}_{\text{phys}}), \quad
    N(\Bsb \to \overline{f}) = N_{\text{avg}} (1 + \mathbb{A}^{\CP}_{\text{phys}}).
    \label{eq:cpv:yield_asym}
\end{equation}
The squared amplitude magnitudes for the individual \Bs and \Bsb decays, $|A_k|^2$ and $|\overline{A}_k|^2$ respectively, which determine the angular shape via the terms $a_k$ (Table~\ref{tab:angAna:decayDescription_SPwave}), are then constructed within the model using the \CP-violation parameter $A^{\CP}_k$, the average magnitudes $|A_k|^2_{\text{avg}}$, and the effective asymmetry $\mathbb{A}^{\CP}_{\text{phys}}$ via the relations
\begin{equation}
    |A_k|^2 = \frac{|A_k|^2_{\text{avg}}(1 - A^{\CP}_k)}{1 - \mathbb{A}^{\CP}_{\text{phys}}}, \quad
    |\overline{A}_k|^2 = \frac{|A_k|^2_{\text{avg}}(1 + A^{\CP}_k)}{1 + \mathbb{A}^{\CP}_{\text{phys}}}.
    \label{eq:cpv:def:N_ACP}
\end{equation}
This parameterisation provides the model for the underlying physics decay rates and shapes based on the parameters ($A^{\CP}_k$, $f_k$, $F_S$, $\delta_k$, $N_{\text{avg}}$) determined by the fit.

% 3. Introduce Raw Asymmetry and Nuisance Effects
However, the experimentally observed yields $N(\Bs)$ and $N(\Bsb)$ are affected by nuisance asymmetries arising from particle production and detection. The raw asymmetry observed in the data is defined as
\begin{equation}
    A^{\rm raw} = \frac {N(\Bsb) - N(\Bs) }
                       {N(\Bsb) + N(\Bs)}.
\label{eq:acp:rawACP_revised} 
\end{equation}
This raw asymmetry differs from the effective physics asymmetry $\mathbb{A}^{\CP}_{\text{phys}}$ (derived from the fit parameters $A^{\CP}_k$ and $|A_k|^2_{\text{avg}}$).
To account for this difference, the likelihood function incorporates the known nuisance asymmetries as fixed inputs.
To do this, when modelling the observed asymmetry between the \Bs and \Bsb samples, the likelihood function uses an effective asymmetry constructed from the physics fit parameters $A^{\CP}_{k}$ combined with the relevant nuisance terms: $\kappa_s A_{\rm P} + A_{\rm D}(K\pi) + A_{\rm PID}(K\pi)$.
Here, $A_{\rm P}$ is the production asymmetry between $\Bsb$ and $\Bs$, 
$A_{\rm D}(K\pi)$ is the detection asymmetry between the $\Kp\pim$ and $\Km\pip$ pairs
%for the charged kaon and pion 
in the final state, 
and $A_{\rm PID}(K\pi)$ is the corresponding asymmetry induced by the hadron PID selection requirements. 
The factor $\kappa_s$ accounts for the dilution due to $B$-meson mixing, following the definition given in Ref.~\cite{LHCb-PAPER-2015-034}.
This procedure allows the fit to determine the true physics asymmetries $A^{\CP}_{k}$.

The production asymmetries $A_{\rm P}(\Bs)$ and $A_{\rm P}(\Bd)$ used in this analysis are taken from measurements performed at $\sqrt{s}=8\tev$~\cite{LHCb-PAPER-2016-062}.
For the \Bs analysis, while the $A_{\rm P}(\Bs)$ value might slightly differ from the true asymmetry at $\sqrt{s}=13\tev$, its impact on the determination of $A^{\CP}_k$ is negligible owing to the very small \Bs mixing dilution factor, $\kappa_s$, determined to be approximately $0.08\%$ for this decay. 
The instrumental detection asymmetry $A_{\rm D}(K\pi)$ and PID asymmetry $A_{\rm PID}(K\pi)$ are evaluated for the Run~2 dataset using dedicated calibration samples in data~\cite{LHCb-PUB-2018-004, LHCb-PUB-2016-021, LHCb-DP-2018-001}.
The calculated asymmetries for the \Bs sample (relevant for the $K^-\pi^+$ final state) are
\begin{align*}
    A_{\text{D}}^{\Bs}(K\pi)    &= (+0.802               \pm 0.030\stat              \pm 0.011\syst) \%, \\
    A_{\text{PID}}^{\Bs}(K\pi)  &= (+0.45\phantom{0}     \pm 0.22\phantom{0}\stat    \pm 0.05\phantom{0}\syst) \%.
\end{align*}
These values are incorporated as fixed inputs into the likelihood fit to extract the physics asymmetries $A^{\CP}_k$. 

The \Bd control channel is analysed analogously. For this channel, the $A_{\rm P}(\Bd)$ value from the $8\tev$ measurement, along with its respective detection and PID asymmetries evaluated for Run~2, are used. The dilution factor for the \Bd channel, $\kappa_d$, is significantly larger (approximately $40\%$); however, the potential discrepancy in $A_{\rm P}(\Bd)$ due to the different centre-of-mass energy is considered small when compared to the statistical precision of the control channel results and the uncertainty on the $A_{\rm P}(\Bd)$ measurement itself, and is thus deemed valid and used in this analysis. 
The results of the fit to the \Bd decay are checked for consistency with \CP symmetry, validating the treatment of the instrumental effects.

% \FloatBarrier
% \newpage

%\FloatBarrier
%\newpage
\section{Branching ratio measurement}
\label{sec:BranchingRatioMeasurement}

The branching fraction of the \mbox{\Bs $\to$ \jpsi\Kstarzb} decay is measured relative to the topologically similar and CKM-favoured decay \mbox{\Bd $\to$ \jpsi\Kstarz}, which acts as a normalisation channel. This relative measurement strategy, identical to that used in Ref.~\cite{LHCb-PAPER-2015-034}, allows for the cancellation of several systematic uncertainties.

\subsection{Measurement strategy}
\label{subsec:BRmeasurement:strategy}

The ratio of branching fractions is given by
\begin{equation}
    \frac{\BRof{\Bs \to \jpsi\Kstarzb}}{\BRof{\Bd \to \jpsi\Kstarz}} =
    \frac{N_{\Bs\to\Kstarzb}}{N_{\Bd\to\Kstarz}} \times \frac{\varepsilon_{\Bd}}{\varepsilon_{\Bs}} \times \frac{f_d}{f_s},
    \label{eq:relBr:basic}
\end{equation}
where $N_{\Bs\to\Kstarzb}$ $(N_{\Bd\to\Kstarz})$ is the yield of signal (normalisation) decays originating purely from  the \pwave $\Kstarzb(\Kstarz)$ resonance, $\varepsilon_{\Bs}$ ($\varepsilon_{\Bd}$) is the corresponding total efficiency, and $f_d/f_s$ is the ratio of fragmentation fractions~\cite{LHCb-PAPER-2020-046}.

The yields $N_{\Bs(\Bd)}$ obtained from the mass fit (Sec.~\ref{sec:FitToTheMassDistribution}) include contributions from all partial waves. Similarly, the efficiencies $\varepsilon_{\Bs(\Bd)}^{\prime}$ estimated from simulation are based on the generator-level angular distributions. Corrections are therefore required, following the method detailed in Ref.~\cite{LHCb-PAPER-2015-034}. 
A factor $F^{\rm P}_{\rm decay}$ corrects the fitted yield to obtain the \pwave yield ($N_{B\to X} = N_{B} \times F^{\rm P}_{\rm decay}$), 
and a factor $c_{\rm decay}$ corrects the simulation-based efficiency to account for the differences between the true angular distribution of the data and that used in the simulation.
Thus, the overall efficiency is \mbox{$\varepsilon = \varepsilon^{\prime} \times c_{\rm decay}$}, where $\varepsilon^{\prime}$ accounts for detector acceptance and reconstruction efficiencies, and $c_{\rm decay}$ accounts for angular mismodelling.
Combining these, the relative branching fraction becomes
\begin{equation}
    \frac{\BRof{\Bs \to \jpsi\Kstarzb}}{\BRof{\Bd \to \jpsi\Kstarz}} =
    \frac{N_{\Bs}}{N_{\Bd}} \times
    \frac{\varepsilon_{\Bd}^{\prime}}{\varepsilon_{\Bs}^{\prime}} \times
    \frac{\zeta_{\Bd}}{\zeta_{\Bs}} \times
    \frac{f_d}{f_s},
    \label{eq:relBr:def:zeta_final}
\end{equation}
where the combined angular correction factor is $\zeta \equiv c_{\rm decay} / F^{\rm P}_{\rm decay}$. The explicit calculation of $F^{\rm P}_{\rm decay}$ and $c_{\rm decay}$ using results from the angular analysis is detailed below. 
The absolute branching fraction for \mbox{\Bs $\to$ \jpsi\Kstarzb} is subsequently obtained using the result from the \belle measurement~\cite{Belle:2002otd}
for \mbox{$\BRof{\Bd \to \jpsi\Kstarz}$}, detailed in Sec.~\ref{subsec:Results_and_uncertainties:Branching_fraction_results}.

\subsection{Relative efficiencies from simulation}
\label{subsec:BranchingRatioMeasurement:Efficiencies_obtained_from_simulation}

The ratio of total efficiencies, \mbox{$\varepsilon_{\Bd}^{\prime} / \varepsilon_{\Bs}^{\prime}$}, accounting for geometrical acceptance as well as trigger, reconstruction, and selection efficiencies, is determined using the corrected simulation samples described in Sec.~\ref{sec:LHCbDetectorAndSimulation}. 
It is determined separately for each data-taking period and then averaged, weighted by the measured signal yields per period. The average efficiency ratio obtained is
\begin{equation*}
    \frac{\varepsilon_{\Bd}^{\prime}}{\varepsilon_{\Bs}^{\prime}} = 0.943 \pm 0.006,
\end{equation*}
where the uncertainty is statistical only.

\subsection{Corrections from angular analysis}
\label{subsec:BranchingRatioMeasurement:Corrections_to_yields_and_efficiencies}

The angular correction factors depend on the measured transversity amplitudes and acceptance weights $\xi_k$. 
Following the methodology described in Ref.~\cite{LHCb-PAPER-2015-034}, the fraction of the fitted yield corresponding to the \pwave component is given by
\begin{equation}
    F^{\rm P}_{\rm{decay}} = \frac{\sum_{k=1}^{6} \xi_k a_k}{\sum_{k=1}^{10} \xi_k a_k C_{k}},
    \label{eq:BR:corr:F}
\end{equation}
and the efficiency correction factor, accounting for differences between the angular distribution measured in data and that used in the simulation, is
\begin{equation}
    c_{\rm decay} = \frac{\sum_{k=1}^{6} \xi_k a_k(\vec{A}_{\rm data})}{\sum_{k=1}^{6} \xi_k a_k(\vec{A}_{\rm sim})}.
    \label{eq:BR:corr:c}
\end{equation}
In these expressions, $a_k$ are the amplitude products (listed in Table~\ref{tab:angAna:decayDescription_SPwave}) evaluated using the measured amplitudes, and $C_k$ incorporates the interference factors, with $C_k=1$ for diagonal terms and $C_k=C_{\rm{SP}}$ for S-P interference. 
For the efficiency correction factor $c_{\rm decay}$, the numerator is evaluated using the amplitudes measured from data ($\vec{A}_{\rm data}$), while the denominator uses the amplitudes assumed in the simulation ($\vec{A}_{\rm sim}$).
The acceptance weights $\xi_k$ are determined once from the simulation sample and are the same in both the numerator and denominator.

These factors are calculated separately for the \Bs and \Bd channels using their respective angular fit results. The ratio of the combined factors required for Eq.~\ref{eq:relBr:def:zeta_final} is determined as
\begin{equation*}
     \frac{\zeta_{\Bd}}{\zeta_{\Bs}} = 1.015 \pm 0.034 \stat \pm 0.017 \syst.
     % \label{eq:BR:corr:zeta_ratio:result}
\end{equation*}
The first uncertainty is statistical, propagated from the angular fit uncertainties, and the second is systematic, obtained by propagating the systematic variations of the angular parameters evaluated in Sec.~\ref{subsec:Results_and_uncertainties:AngularParaAndCPPara}.
With all components for the calculation of the relative branching fraction (Eq.~\ref{eq:relBr:def:zeta_final}) determined, the results are presented in Sec.~\ref{subsec:Results_and_uncertainties:Branching_fraction_results}.

%\FloatBarrier
%\newpage
\section{Results and uncertainties}
\label{sec:ResultsAndUncertainties}

This section presents the results of the angular analysis and the branching fraction measurement based on the Run~2 dataset. The determination of statistical and systematic uncertainties follows standard procedures, similar to those described in the Run~1 analysis~\cite{LHCb-PAPER-2015-034}, adapted for the specific conditions and models used here.

\subsection{Angular and \texorpdfstring{\boldmath $\CP$}{}-violation parameters}
\label{subsec:Results_and_uncertainties:AngularParaAndCPPara}

\begin{figure}[!tb]
     \includegraphics[scale=0.39]{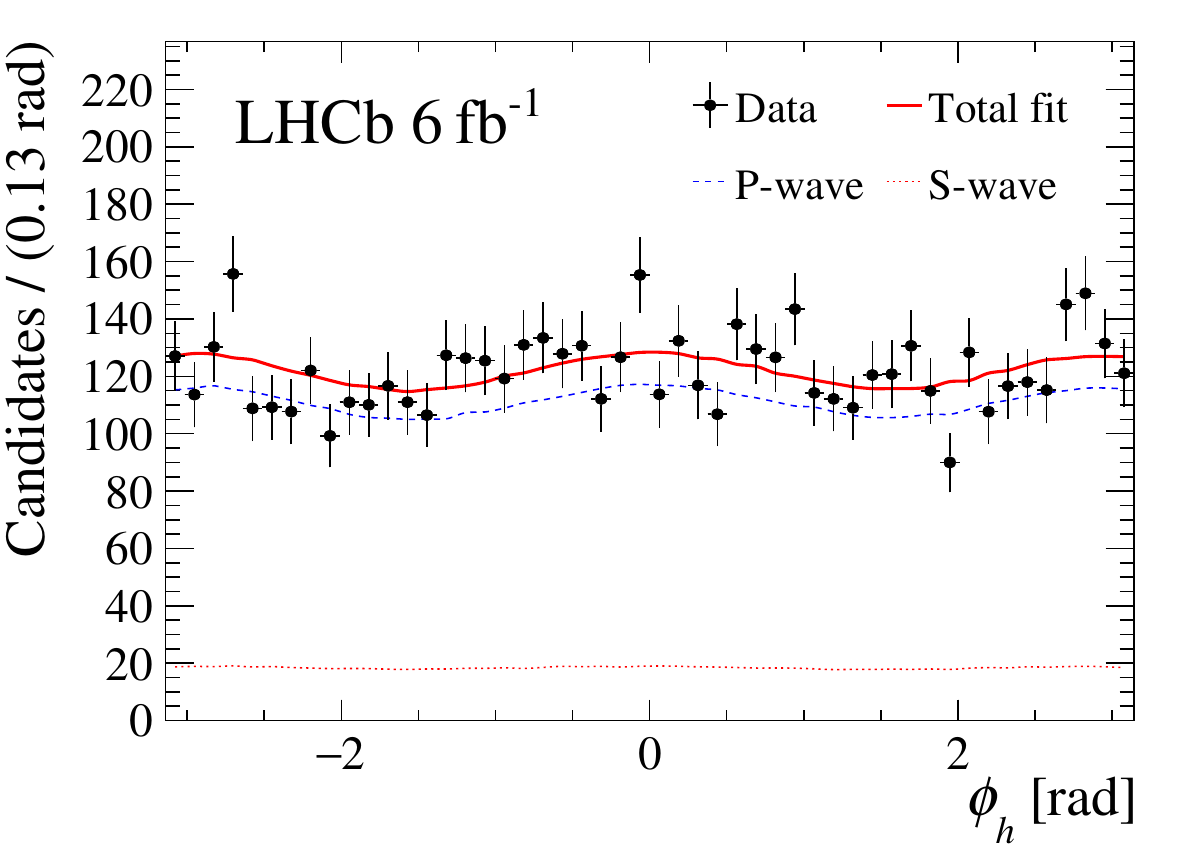}
     \includegraphics[scale=0.39]{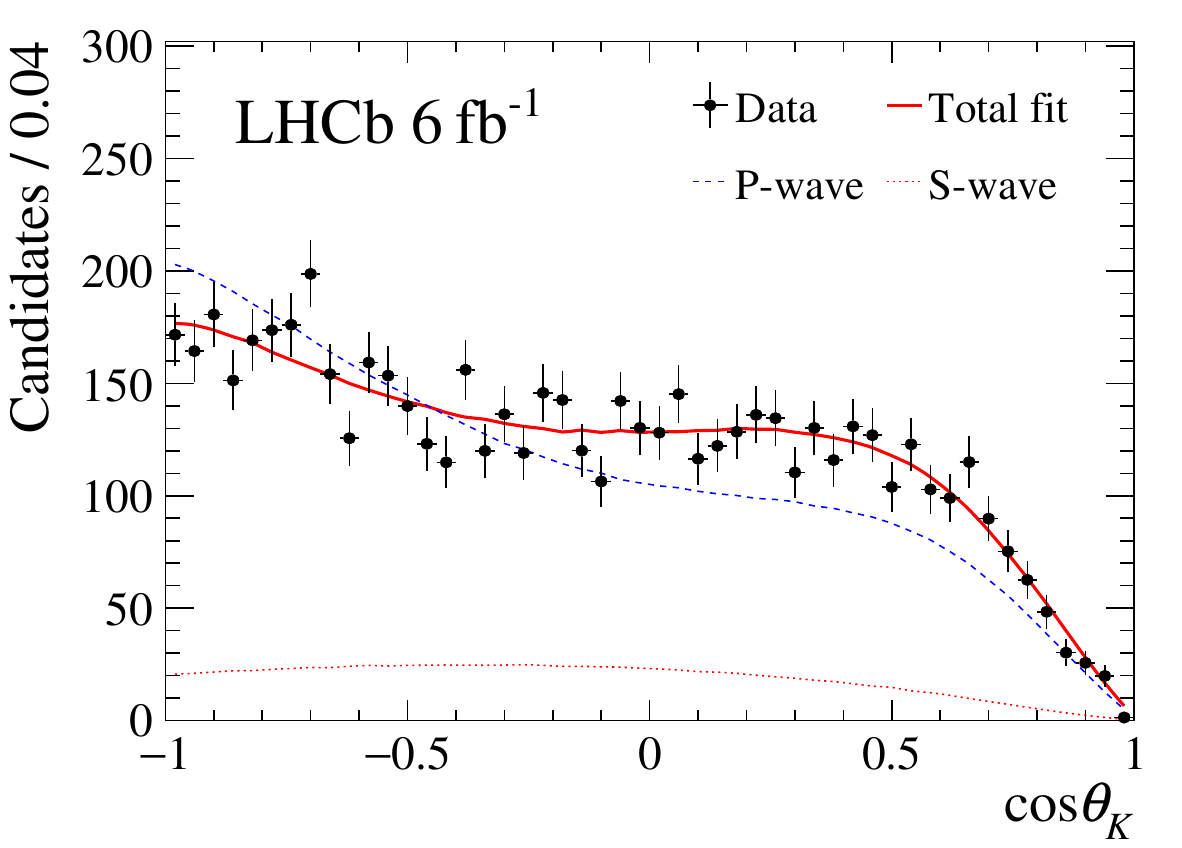}\\
     \centering
     \includegraphics[scale=0.39]{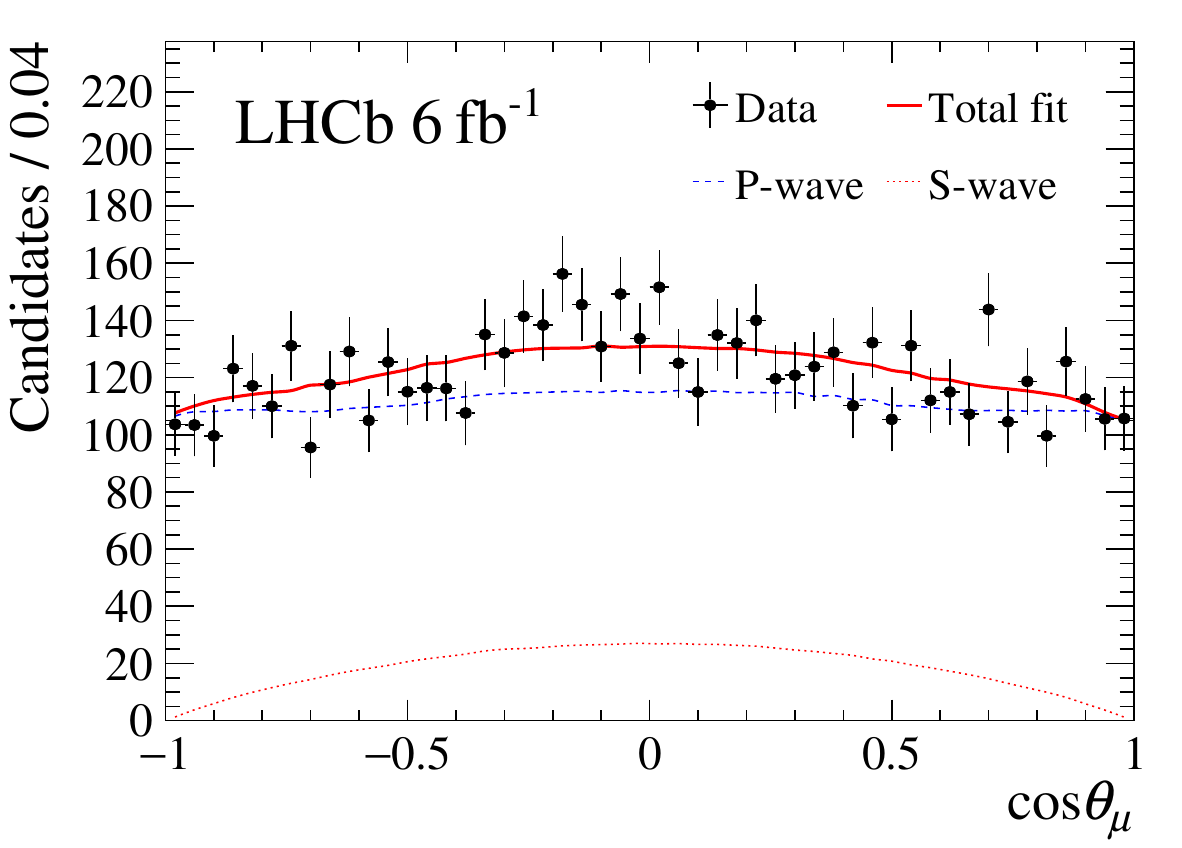}
     \caption{
         Projections of the angular fit result onto the helicity angles 
         (top left) $\cos{\theta_{K}}$, 
         (top right) $\cos{\theta_{\mu}}$, 
         and (bottom) $\phi_{h}$ 
         for the background-subtracted 
         $B^0_s  \rightarrow  J\mskip -3mu/\mskip -2mu\psi  \Kbar{}^{*}\kern-1pt(892)^{0}$ data sample (black points with error bars), summed over all subsamples. 
        The overall fit model (red solid line) is shown, along with its \text{P-wave} (blue dashed) and \text{S-wave} (orange dotted) components. 
        Interference terms between S- and \text{P-wave}, which are part of the total fit model, are not explicitly shown but contribute to the overall shape; their effect can lead to regions where individual wave components appear to exceed the total fit projection.
     }
     \label{fig:angAna:final:cmp:Bs}
\end{figure}

The central values for the \pwave and \swave parameters are determined from the simultaneous angular fit described in Sec.~\ref{sec:AngularAnalysis}. 
The main results for the \pwave parameters from the Run~2 dataset are presented below.
Detailed results for the \swave parameters, determined per \mkpi bin, and the full breakdown of systematic uncertainties for all parameters, are given in Tables~\ref{tab:finalResults:overall:Bs:Pwave} and \ref{tab:finalResults:overall:Bs:Swave}.
Figure~\ref{fig:angAna:final:cmp:Bs} shows the projection of the fit onto the background-subtracted data. 
The full correlation matrix for the fitted parameters is provided in Appendix~\ref{subapp:AngularAna:Correlation_matrix}.

\subsubsection{Run~2 \pwave polarisation fractions and \texorpdfstring{\boldmath $\CP$}{} asymmetries} % New subsection
\label{subsubsec:Run2PwaveResults}

The \pwave polarisation fractions obtained from the Run~2 data, averaged over the \mkpi range, are
\[
  \setlength{\arraycolsep}{1mm}
  \begin{array}{lclllllll}
        \fL              &\;=\; & 0.534   &\pm & 0.012\stat  &\pm & 0.009\syst, \\
        \fpa             &\;=\; & 0.211   &\pm & 0.014\stat  &\pm & 0.005\syst.
  \end{array}
\]
The corresponding direct \CP asymmetries for the \pwave components, also averaged over the \mkpi range, are measured to be
\[
  \setlength{\arraycolsep}{1mm}
  \begin{array}{lclllllll}
        \ACPL           &\;=\; & \phantom{-}0.014   &\pm & 0.029\stat  &\pm & 0.007\syst, \\
        \ACPpa          &\;=\; &          -0.055  &\pm & 0.065\stat  &\pm & 0.007\syst, \\
        \ACPpe          &\;=\; & \phantom{-}0.060   &\pm & 0.057\stat  &\pm & 0.016\syst.
  \end{array}
\]
These \CP asymmetries are consistent with zero. The phases $\delta_{\parallel}$ and $\delta_{\perp}$ are also determined and presented in Table~\ref{tab:finalResults:overall:Bs:Pwave}.

% Figure placement

Statistical uncertainties are determined using a bootstrap method~\cite{efron:1979}, where the entire analysis chain is repeated on multiple datasets resampled from the original data.
Systematic uncertainties are obtained by considering variations in the analysis procedure, modelling assumptions, and external inputs.
Individual systematic uncertainties are assumed to be uncorrelated,
and the total systematic uncertainty is calculated as the quadratic sum of these contributions. 
The main sources considered are summarised below, grouped according to their origin. Detailed contributions for each parameter are listed in Tables~\ref{tab:finalResults:overall:Bs:Pwave} and \ref{tab:finalResults:overall:Bs:Swave}.

\subsubsection{Systematic uncertainties related to the mass fit}
\label{subsec:Results_and_uncertainties:AngularPara_and_CPPara:Systematics_related_to_massfit}
These uncertainties affect the calculation of the \sPlot weights and consequently propagate to the angular analysis, influencing the measurement of angular parameters.
\begin{description}
    \item[Mass fit model:] Assessed by repeating the analysis using alternative PDF shapes (Hypatia function for signal peaks~\cite{Santos:2013gra}, Chebyshev polynomial for background) for the mass fit (Sec.~\ref{sec:FitToTheMassDistribution}).
    \item[Peaking backgrounds:] Determined by varying the yields of the statistically subtracted peaking backgrounds (Sec.~\ref{sec:PeakingBackground}) within their uncertainties ($\pm 3\sigma$).

    \item[Exotic contributions:] The potential impact of exotic state contributions, such as $T_{\cquark\cquarkbar1}(4200)^{\pm}$ and $T_{\cquark\cquarkbar1}(4430)^{\pm}$ produced in $B \to T_{\cquark\cquarkbar1} K$ ($T_{\cquark\cquarkbar1} \to \jpsi \pi$) decays~\cite{PhysRevD.90.112009}, which yield the same final-state particles as the signal, is estimated using the \Bd control channel. An additional selection $\costk > -0.6$ is applied to the \Bd sample, designed empirically to suppress such contributions. The systematic uncertainty assigned to the corresponding \Bs analysis results is taken as the maximum deviation observed in the \Bd parameters due to this selection.

    \item \textbf{Mass--angle correlations:} Potential biases from residual correlations between \mbox{$m(\jpsi\Kmpip)$} and helicity angles are primarily evaluated by performing the mass fit in five bins of \costmu. Correlations with the other helicity angles, \costk and \phihel, are found to be negligible, and their potential effects are neglected.

\end{description}

\FloatBarrier
%\newpage
\subsubsection{Systematic uncertainties related to the angular analysis}
\label{subsec:Results_and_uncertainties:AngularPara_and_CPPara:Systematics_only_angular}
These uncertainties arise from the angular fitting procedure, the inputs to the fit, and the modelling of the angular distributions.
They directly impact the determination of the polarisation fractions, phase differences, and \CP asymmetries.
\begin{description}
    \item[Simulation sample size] Estimated by bootstrapping the simulation samples used to calculate the acceptance normalisation weights ($\xi_k$).
    
    \item[\boldmath {$C_{SP}$} factor] Assessed by varying the $C_{\rm{SP}}$ values based on different lineshape models for the interfering S- and P-waves, following the procedure in Ref.~\cite{LHCb-PAPER-2015-034}.
    
    \item[\dwave contribution:] Estimated using pseudoexperiments generated with a small \dwave component (found to be \mbox{$\sim 0.4\%$} on average from a fit to data) and fitted with the baseline model.
    
    \item[\boldmath \CP violation in $\delta_k$ phases:] 
    % \item[\boldmath \mbox{$\Delta\delta_k \neq 0$}:] 
    Evaluated by allowing for potential differences between the $\delta_k$ phases for \Bs and \Bsb decays (\mbox{$\Delta\delta_k \equiv \delta_k(\Bs) - \delta_k(\Bsb) \neq 0$}) for the \pwave components in a variant fit.
    
    \item[Simulation corrections:] Assessed by comparing results using the baseline simulation corrections (Sec.~\ref{subsec:AngularAcceptance}, kinematic reweighting primarily from \Bd mesons) versus corrections derived directly using only \Bs samples.
    
    \item[Fit model bias:] Checked using pseudoexperiments generated according to the baseline fit result and fitted with the same procedure.
    
    \item[\boldmath External \texorpdfstring{\CP}{} asymmetries:] Uncertainties from the measured production and instrumental asymmetries ($A_{\rm P}, A_{\rm D}, A_{\rm PID}$) are propagated to the physics \CP asymmetries $A^{\CP}_k$.
\end{description}

%%%%%%%%%%%%%%%%%%%%%%%%%%%%%%%%%%%%%%%%%%%%%%%%%%%%%%%%%%%%%%%%%%%%%%%%%%%%%%%%%%%%%%%%%%%%%%%%
%%%%%%%%%%%%%%%%%%%%%%%%%%%%%%%%        P wave      %%%%%%%%%%%%%%%%%%%%%%%%%%%%%%%%
\begin{table}[htbp]
    \begin{center}
    \caption{Summary of results and uncertainties for \pwave parameters obtained in \mbox{\BsJpsiKst} decays, from the Run~2 data analysis.
    Central values and statistical uncertainties are from the baseline fit. Systematic uncertainties from different sources are listed, along with their quadratic sum. The total uncertainty is the quadratic sum of statistical and total systematic uncertainties. The phases are shown in radians. Values below $5\times10^{-4}$ are indicated by ``---''.}
    \label{tab:finalResults:overall:Bs:Pwave}
    % \vspace{5mm} % Add space for placeholder content
    \centering

    \resizebox{\textwidth}{!}{%
        {\renewcommand{\arraystretch}{1.8} % Apply line spacing only to this table
        
            \begin{tabular}{lcccccccc}

\hline
\hline

     Parameter &   \ACPL &  \ACPpa &  \ACPpe &  $\fL$ &  $\fpa$ &  $\delta_{\parallel}$ &   $\delta_{\perp}$
     \\ 
     % [+0.04in]
    \hline

 Measured value                   &  $0.014$      &  $-0.055$     &  $0.060$      &  $0.534$      &  $0.211$      &  $2.879$      &  $0.057$        \\   \hline

 Statistical uncertainty          &  $0.029$      &  $0.065$      &  $0.057$      &  $0.012$      &  $0.014$      &  $0.085$      &  $0.065$        \\   \hline

    % ======================    mass fit   ======================

    %
 Mass fit model                   &  $0.001$      &  ---          &  ---          &  $0.001$      &  ---          &  ---          &  $0.001$        \\ 
 Peaking bkg.                     &  ---          &  $0.001$      &  ---          &  ---          &  ---          &  $0.002$      &  $0.001$        \\ 
 Exotics pollution                &  $0.003$      &  $0.001$      &  $0.003$      &  $0.004$      &  $0.002$      &  $0.012$      &  $0.012$        \\ 
 Mass--\costmu correlation       &  $0.002$      &  ---          &  $0.004$      &  $0.002$      &  $0.001$      &  $0.005$      &  $0.002$        \\ 
 Ang. Acc. (Sim. size)             &  $0.001$      &  $0.004$      &  $0.004$      &  $0.001$      &  $0.001$      &  $0.005$      &  $0.004$        \\ 
 $C_{\rm{SP}}$ factor                  &  ---          &  $0.002$      &  $0.001$      &  ---          &  $0.001$      &  $0.003$      &  $0.005$        \\ 
 \dwave contribution        &  $0.005$      &  $0.002$      &  $0.014$      &  $0.001$      &  ---          &  $0.001$      &  $0.003$        \\ 
 % \CPV in phase differences
     \mbox{$\Delta\delta_k \neq 0$}
    &  ---          &  ---          &  ---          &  ---          &  ---          &  $0.004$      &  $0.003$        \\ 
 Simulation corrections         &  ---          &  $0.004$      &  ---          &  $0.008$      &  $0.004$      &  $0.009$      &  $0.010$        \\ 
 Bias in model                    &  $0.001$      &  $0.001$      &  ---          &  $0.001$      &  $0.001$      &  $0.001$      &  $0.001$        \\ 
    
 External \texorpdfstring{\CP}{} asym. 
& $0.002$ &  $0.002$ &  $0.002$
& --- &  --- &  --- &  ---  
 \\
\hline

 Quadratic sum of syst.           &  $0.007$      &  $0.007$      &  $0.016$      &  $0.009$      &  $0.005$      &  $0.018$      &  $0.018$        \\   \hline

% ======================    total uncertainties   ======================
 Total uncertainty                
&  $0.030$      &  $0.066$      &  $0.059$      
&  $0.015$      &  $0.015$      &  $0.087$      & $0.068$        \\

\hline
\hline
            \end{tabular}
        } % resize
    } % \renewcommand

    \end{center}
\end{table}

%%%%%%%%%%%%%%%%%%%%%%%%%%%%%%%%%%%%%%%%%%%%%%%%%%%%%%%%%%%%%%%%%%%%%%%%%%%%%%%%%%%%%%%%%%%%%%%%
%%%%%%%%%%%%%%%%%%%%%%%%%%%%%%%%        S wave      %%%%%%%%%%%%%%%%%%%%%%%%%%%%%%%%
\begin{table}[htbp]
\begin{center}
\caption{Summary of results and uncertainties for \swave parameters obtained in \mbox{$\Bs \to$ \jpsi\Km\pip} decays, from the Run~2 data analysis. 
Central values and statistical uncertainties are from the baseline fit. Systematic uncertainties from different sources are listed, along with their quadratic sum. The total uncertainty is the quadratic sum of statistical and total systematic uncertainties. The phases are shown in radians. Values below $5\times10^{-4}$ are indicated by ``---''.
}
\label{tab:finalResults:overall:Bs:Swave}
% \vspace{5mm} % Add space for placeholder content
\centering

    \resizebox{\textwidth}{!}{%
        {\renewcommand{\arraystretch}{1.8} % Apply line spacing only to this table
        
            % \footnotesize % Adjust font size if needed
            \begin{tabular}{lc|cc|cc|cc|cc}
%\hline
%
\hline
\hline
    \multirow{2}{*}{Parameter}  &  \multirow{2}{*}{\ACPS} & \multicolumn{2}{c|}{$\mkpi^{\rm bin0}$} & \multicolumn{2}{c|}{$\mkpi^{\rm bin1}$} & \multicolumn{2}{c|}{$\mkpi^{\rm bin2}$} & \multicolumn{2}{c}{$\mkpi^{\rm bin3}$} \\ 
                                & & \multicolumn{1}{c}{$F_S$} & \multicolumn{1}{c|}{$\delta_S$} & $F_S$ & \multicolumn{1}{c|}{$\delta_S$} & $F_S$ & \multicolumn{1}{c|}{$\delta_S$} & $F_S$ & $\delta_S$ 
     \\ 
     % [+0.04in]
    \hline

Measured value                   & $0.081$      & $0.420$      & $-0.557$     & $0.076$      & $0.556$      & $0.083$      & $1.376$      & $0.243$      & $1.861$        \\
    
    \hline
Statistical uncertainty          & $0.085$      & $0.096$      & $0.110$      & $0.019$      & $0.161$      & $0.023$      & $0.075$      & $0.057$      & $0.083$        \\
    \hline

    % ======================    mass fit   ======================
Mass fit model                   & $0.002$      & $0.001$      & $0.005$      & ---          & $0.003$      & $0.001$      & $0.004$      & $0.001$      & $0.002$        \\  
Peaking bkg.                     & $0.001$      & $0.002$      & $0.003$      & ---          & ---          & ---          & $0.003$      & $0.001$      & $0.005$        \\  
Exotics pollution                & $0.001$      & $0.002$      & $0.006$      & ---          & $0.008$      & $0.001$      & $0.005$      & $0.007$      & $0.019$        \\  
Mass--\costmu correlation       & $0.015$      & $0.042$      & $0.037$      & $0.001$      & $0.003$      & $0.006$      & $0.003$      & $0.008$      & $0.004$        \\  
Ang. Acc. (Sim. size)             & $0.007$      & $0.008$      & $0.008$      & $0.001$      & $0.012$      & $0.002$      & $0.005$      & $0.004$      & $0.006$        \\  
$C_{\rm{SP}}$ factor                  & $0.001$      & $0.012$      & $0.052$      & $0.004$      & $0.001$      & $0.003$      & $0.001$      & $0.003$      & $0.006$        \\  
\dwave contribution        & $0.018$      & $0.001$      & $0.016$      & $0.002$      & $0.002$      & $0.002$      & $0.001$      & $0.002$      & $0.001$        \\  
% \CPV in phase differences
     \mbox{$\Delta\delta_k \neq 0$}
& $0.001$      & $0.002$      & $0.001$      & ---          & $0.002$      & ---          & ---          & $0.003$      & $0.002$        \\  
Simulation corrections         & ---          & $0.051$      & $0.011$      & $0.002$      & $0.003$      & $0.002$      & $0.009$      & $0.018$      & $0.060$        \\  
Bias in model                    & $0.002$      & $0.003$      & $0.017$      & $0.003$      & $0.006$      & $0.001$      & $0.003$      & $0.001$      & $0.003$        \\

External \texorpdfstring{\CP}{} asym.
& $0.002$ 
& --- & --- & --- & ---
& --- & --- & --- & --- \\    \hline

Quadratic sum of syst.           & $0.024$      & $0.068$      & $0.070$      & $0.006$      & $0.017$      & $0.007$      & $0.013$      & $0.022$      & $0.064$        \\  \hline

% ======================    total uncertainties   ======================
Total uncertainty                & $0.088$      & $0.118$      & $0.131$      & $0.020$      & $0.162$      & $0.024$      & $0.076$      & $0.061$      & $0.104$       \\

\hline
\hline

            \end{tabular}
        } % resize
    } % \renewcommand

\end{center}
\end{table}

%\FloatBarrier 
\subsection{Branching fraction}
\label{subsec:Results_and_uncertainties:Branching_fraction_results}

Substituting the ratio of fitted yields $N_{\Bs}/N_{\Bd}$ (Sec.~\ref{sec:FitToTheMassDistribution}), the efficiency ratio $\varepsilon_{\Bd}^{\prime} / \varepsilon_{\Bs}^{\prime}$ (Sec.~\ref{subsec:BranchingRatioMeasurement:Efficiencies_obtained_from_simulation}), the angular correction ratio $\zeta_{\Bd}/\zeta_{\Bs}$ (Sec.~\ref{subsec:BranchingRatioMeasurement:Corrections_to_yields_and_efficiencies}), and the hadronisation fraction ratio \mbox{$f_d/f_s = 3.939 \pm 0.123$}~\cite{LHCb-PAPER-2020-046} into Eq.~\ref{eq:relBr:def:zeta_final}, the relative branching fraction for the Run~2 dataset is found to be
\begin{align*}
    \frac{\BRof{\Bs \to \jpsi\Kstarzb}}{\BRof{\Bd \to \jpsi\Kstarz}}
    &= (3.08\pm 0.11 \stat \pm 0.06 \syst  \pm 0.10 \,(\fdfs))\% .
    % \label{eq:BR:rel_results:Run2} % Kept original label
\end{align*}
The statistical uncertainty combines contributions from the yields and the ratio of the angular correction factors. The systematic uncertainty includes contributions from the fit model variations and the angular-correction ratio. The uncertainty labelled $\fdfs$ is due to the hadronisation fraction ratio.
The dominant systematic uncertainties on the relative branching fraction originate from the choice of mass fit models (Sec.~\ref{subsec:Results_and_uncertainties:AngularPara_and_CPPara:Systematics_related_to_massfit}) and the propagation of uncertainties from the angular analysis parameters (evaluated in Sec.~\ref{subsec:Results_and_uncertainties:AngularParaAndCPPara}) via the correction factor ratio $\zeta_{\Bd}/\zeta_{\Bs}$ (defined in Sec.~\ref{subsec:BranchingRatioMeasurement:Corrections_to_yields_and_efficiencies}).
Other potential systematic uncertainties are found to largely cancel in the ratio.

The absolute branching fraction \mbox{$\BRof{\Bs \to \jpsi\Kstarzb}$} from the Run~2 dataset is obtained by multiplying the relative branching fraction measured above by an external value for the branching fraction of the normalisation channel, \mbox{$\BRof{\Bd \to \jpsi\Kstarz}$}. 
For consistency with the amplitude analysis approach used here and in the previous \lhcb measurement~\cite{LHCb-PAPER-2015-034}, which explicitly separates partial waves but without explicit consideration of exotic states such as $T_{\cquark\cquarkbar1}$ resonances, the result from the \belle collaboration~\cite{Belle:2002otd} is used instead of the PDG average~\cite{PDG2024},
\mbox{$
    \BRof{\Bd \to \jpsi \Kstarz}_{\rm Belle} = (1.29 \pm 0.05 \stat \pm 0.13 \syst) \times 10^{-3}
$}.
As this measurement was performed at the $\FourS$ resonance assuming equal production fractions $f^{00}=f^{+-}=0.5$, 
a correction is needed because the measured production ratio $R^{+/0} \equiv f^{+-}/f^{00} = \Gamma(\FourS \to \Bu\Bub)/\Gamma(\FourS \to \Bd\Bdb) = 1.052 \pm 0.031$~\cite{HFLAV23} deviates from unity.
The correction factor is calculated as \mbox{$C = (1+R^{+/0})/2 = 1.026 \pm 0.016$}.

Using this correction factor, the absolute branching fraction for the signal decay from the Run~2 dataset is determined as
\begin{align*}
    \BRof{\Bs \to \jpsi\Kstarzb}
    &= \frac{\BRof{\Bs \to \jpsi\Kstarzb}}{\BRof{\Bd \to \jpsi\Kstarz}} \times C \times \BRof{\Bd \to \jpsi \Kstarz}_{\rm Belle} \nonumber \\
    &= (4.07 \pm 0.15 \stat \pm 0.07 \syst \pm 0.13 \,(\fdfs) \pm 0.45 \,(\BR_{\Bd})) \times 10^{-5}.
    % \label{eq:BR:abs_result:Run2}
\end{align*}
The uncertainties are ordered as: statistical, systematic, from $f_d/f_s$, and from other external inputs.
This result is consistent with the previous \lhcb measurement~\cite{LHCb-PAPER-2015-034} and the current world average~\cite{PDG2024}.

%\FloatBarrier
%\newpage
\section{Combination with \lhcb Run~1 results}
\label{sec:CombinationWithLHCbRun1Results}

The results obtained in this analysis (Sec.~\ref{sec:ResultsAndUncertainties}) are consistent with the previous \lhcb measurements using the Run~1 dataset~\cite{LHCb-PAPER-2015-034}. A combination of the Run~1 and Run~2 results is performed using the Best Linear Unbiased Estimator (BLUE) method~\cite{Nisius:2020jmf}, taking into account correlations between uncertainties.

For the angular parameters, statistical uncertainties are treated as uncorrelated between the two datasets. 
Systematic uncertainties related to shared theoretical inputs or methodologies (\eg $C_{\rm{SP}}$ factor modelling, external asymmetry evaluations) are considered 100\% correlated, while dataset-specific uncertainties (\eg simulation sample size, fit model choices specific to Run~1 or Run~2, background modelling) are treated as uncorrelated. The combined results for the \pwave polarisation fractions and \CP asymmetries are
  \[
  \setlength{\arraycolsep}{1mm}
  \begin{array}{cclllllll}
        \fL             &\;=\; & \phantom{-}0.528   &\pm & 0.011 \stat  &\pm & 0.009 \syst ,  \\
        \fpa            &\;=\; & \phantom{-}0.205   &\pm & 0.012 \stat  &\pm & 0.005 \syst ,    \\
%       \fpe            &\;=\; &        xx          &\pm & xx   \stat   &\pm & xxx \syst    \\
        \ACPL       &\;=\; & \phantom{-}0.021   &\pm & 0.026 \stat  &\pm & 0.007 \syst ,    \\
        \ACPpa      &\;=\; &          - 0.073   &\pm & 0.060 \stat  &\pm & 0.007 \syst ,    \\
        \ACPpe   &\;=\; & \phantom{-}0.057   &\pm & 0.049 \stat  &\pm & 0.014 \syst .\\
  \end{array}
  \]

\FloatBarrier
%\newpage

For the branching fraction combination, 
the relative measurement \mbox{$\BRof{\Bs \to \jpsi\Kstarzb} / \BRof{\Bd \to \jpsi\Kstarz}$} from Run~1~\cite{LHCb-PAPER-2015-034} is updated before being used. It was originally reported as \mbox{$(2.99 \pm 0.14 \stat \pm 0.12 \syst \pm 0.17 \,(\fdfs))\%$}, depended on an older determination of the hadronisation fraction ratio~\cite{fsfd_run1}.
Since the measurement scales with $f_d/f_s$, this Run~1 result is recalculated using the latest value appropriate for Run~1 conditions, \mbox{$f_s/f_d = 0.2387 \pm 0.0075$}~\cite{LHCb-PAPER-2020-046}. This recalculation yields an updated Run~1 branching fraction ratio of \mbox{$(3.24 \pm 0.15 \stat \pm 0.13 \syst \pm 0.11 \,(\fdfs))\%$}, which is used in the combination presented below.

The updated Run~1 value is combined with the Run~2 result from Sec.~\ref{subsec:Results_and_uncertainties:Branching_fraction_results} using the BLUE method. Statistical uncertainties are treated as uncorrelated. The systematic uncertainty arising from the external $f_d/f_s$ ratio is treated as 100\% correlated, while other systematic uncertainties specific to each analysis are assumed uncorrelated. The combined relative branching fraction is found to be
\begin{equation*}
    \frac{\BRof{\Bs \to \jpsi\Kstarzb}}{\BRof{\Bd \to \jpsi\Kstarz}} = (3.12 \pm 0.09 \stat \pm 0.06 \syst \pm 0.10 \,(\fdfs))\% .
    \label{eq:BR:rel_results:combined}
\end{equation*}
The combined absolute branching fraction is obtained by multiplying this ratio by the corrected \belle branching fraction for the normalisation channel (defined in Sec.~\ref{subsec:Results_and_uncertainties:Branching_fraction_results}),
and is determined as
\begin{equation*}
  \BRof{\Bs \to \jpsi\Kstarzb} = (4.13 \pm 0.12 \stat \pm 0.07 \syst \pm 0.14 \,(\fdfs) \pm 0.45 \,(\BR_{\Bd})) \times 10^{-5} .
\end{equation*}

%\FloatBarrier
%\newpage
\section{Conclusions}
\label{sec:Conslusions}

This paper presents an angular analysis of the \mbox{\BsJpsiKstI} decay, performed using \proton\proton collision data collected by the \lhcb experiment at \mbox{$\sqrt{s}=13\tev$} during 2015--2018, corresponding to an integrated luminosity of \mbox{$6\invfb$}. 
The \pwave polarisation fractions and direct \CP asymmetries, presented in Sec.~\ref{subsubsec:Run2PwaveResults}, along with \swave parameters and the branching fraction (Sec.~\ref{subsec:Results_and_uncertainties:Branching_fraction_results}), significantly improve upon previous measurements. The measured \CP asymmetries are consistent with zero within uncertainties.

Combining these Run~2 results with those from Run~1~\cite{LHCb-PAPER-2015-034}, as detailed in Sec.~\ref{sec:CombinationWithLHCbRun1Results}, yields the most precise determinations to date of the angular parameters and the relative branching fraction $\BRof{\BsJpsiKstI}/\BRof{\BdJpsiKstI}$. 
The combination also provides an updated value for the absolute branching fraction, determined using external input for the normalisation channel from the \belle collaboration~\cite{Belle:2002otd}.
These precise results provide crucial inputs for constraining penguin contributions to the \CP-violating phase \phis measured in \mbox{$\bquark \to \ccbar\squark$} decays.
However, as established in the previous analysis of this channel~\cite{LHCb-PAPER-2015-034}, a quantitative constraint based solely on \BsJpsiKst decays remains limited by theoretical uncertainties in the hadronic form factors. 
The full benefit of the improved experimental precision will therefore only be realised in a future combined analysis that also incorporates results from \grpsuthree-partner decays, such as \BdJpsiRho.
Continued improvements in experimental precision with larger datasets at \lhcb are anticipated to further refine these constraints and enhance the sensitivity to potential new physics in the \Bs meson system.

% Do not include this in any draft (just for information in the template)
% \input{acknowledgements_intro}
% Comment this in for paper drafts; do not include this in analysis note, conference and figure reports
\section*{Acknowledgements}
%
% These Acknowledgements valid from 3-May-2019
%
\noindent We express our gratitude to our colleagues in the CERN
accelerator departments for the excellent performance of the LHC. We
thank the technical and administrative staff at the LHCb
institutes.
We acknowledge support from CERN and from the national agencies:
ARC (Australia);
CAPES, CNPq, FAPERJ and FINEP (Brazil); 
MOST and NSFC (China); 
CNRS/IN2P3 (France); 
BMBF, DFG and MPG (Germany); 
INFN (Italy); 
NWO (Netherlands); 
MNiSW and NCN (Poland); 
MCID/IFA (Romania); 
%MSHE (Russia); 
MICIU and AEI (Spain);
SNSF and SER (Switzerland); 
NASU (Ukraine); 
STFC (United Kingdom); 
DOE NP and NSF (USA).
%%%%%%%%%%%%%%%%%%%%%%%%%%%%%%%%%%%%%%%%%%%%%
We acknowledge the computing resources that are provided by ARDC (Australia), 
CBPF (Brazil),
CERN, 
IHEP and LZU (China),
IN2P3 (France), 
KIT and DESY (Germany), 
INFN (Italy), 
SURF (Netherlands),
Polish WLCG (Poland),
IFIN-HH (Romania), 
%RRCKI and Yandex LLC (Russia), 
PIC (Spain), CSCS (Switzerland), 
and GridPP (United Kingdom).
%%%%%%%%%%%%%%%%%%%%%%%%%%%%%%%%%%%%%%%%%%
We are indebted to the communities behind the multiple open-source
software packages on which we depend.
%%%%%%%%%%%%%%%%%%%%%%%%%%%%%%%%%%%%%%%%%%
Individual groups or members have received support from
%ARC and ARDC (Australia); % moved to national 16/01
Key Research Program of Frontier Sciences of CAS, CAS PIFI, CAS CCEPP, 
Fundamental Research Funds for the Central Universities,  and Sci.\ \& Tech.\ Program of Guangzhou (China);
Minciencias (Colombia);
EPLANET, Marie Sk\l{}odowska-Curie Actions, ERC and NextGenerationEU (European Union);
A*MIDEX, ANR, IPhU and Labex P2IO, and R\'{e}gion Auvergne-Rh\^{o}ne-Alpes (France);
%RFBR, RSF and Yandex LLC (Russia);
Alexander-von-Humboldt Foundation (Germany);
ICSC (Italy); 
%GVA, XuntaGal, GENCAT, Inditex, InTalent and Prog.~Atracci\'on Talento, CM (Spain);
Severo Ochoa and Mar\'ia de Maeztu Units of Excellence, GVA, XuntaGal, GENCAT, InTalent-Inditex and Prog.~Atracci\'on Talento CM (Spain);
SRC (Sweden);
the Leverhulme Trust, the Royal Society and UKRI (United Kingdom).

\section*{Appendices}

\appendix

\FloatBarrier
% \newpage
\section{Correlation matrix} 
\label{subapp:AngularAna:Correlation_matrix}

The correlations between model parameters used in the angular analysis of \Bs signal decays are listed in Table.~\ref{tab:angAna:data:Bs:correlationMatrix:SPwave}.

\begin{table}[htbp]
\small
\centering
\caption{
    Statistical correlation matrix for the physics parameters in the angular fit on \BsJpsiKst decays.
    The \swave fraction and phase are allowed to vary freely in each of these \mkpi bins. 
    These bin-dependent parameters are denoted using superscripts indicating the bin range, for example, $\AsBinZero$ and  $\dsBinZero$ represent the \swave fraction and phase, respectively, in the first \mkpi bin covering the range $826$ to $861\mevcc$. 
}

\renewcommand{\arraystretch}{2.0}
\fontsize{8pt}{8pt}\selectfont
\setlength\tabcolsep{0.420cm}
\rotatebox{90}{% <-- Rotate the table

\begin{tabularx}{21.0cm}{
@{}
lp{0.245cm} 
 p{0.245cm} 
 p{0.245cm} 
 p{0.245cm} 
 p{0.245cm} 
 p{0.245cm} 
 p{0.245cm} 
 p{0.245cm} 
 p{0.245cm} 
 p{0.245cm} 
 p{0.245cm} 
 p{0.245cm} 
 p{0.245cm} 
 p{0.245cm} 
 p{0.245cm} 
 p{0.245cm}@{}}

\toprule
\multicolumn{1}{@{}l}{\textbf{}} 
& \multicolumn{1}{c}{\rotatebox[origin=c]{270}{\textbf{\ACPS}}} 
& \multicolumn{1}{c}{\rotatebox[origin=c]{270}{\textbf{\ACPL}}} 
& \multicolumn{1}{c}{\rotatebox[origin=c]{270}{\textbf{\ACPpa}}} 
& \multicolumn{1}{c}{\rotatebox[origin=c]{270}{\textbf{\ACPpe}}} 
& \multicolumn{1}{c}{\rotatebox[origin=c]{270}{\textbf{\AsBinZero}}} 
& \multicolumn{1}{c}{\rotatebox[origin=c]{270}{\textbf{\AsBinOne}}} 
& \multicolumn{1}{c}{\rotatebox[origin=c]{270}{\textbf{\AsBinTwo}}} 
& \multicolumn{1}{c}{\rotatebox[origin=c]{270}{\textbf{\AsBinThree}}} 
& \multicolumn{1}{c}{\rotatebox[origin=c]{270}{\textbf{\dsBinZero}}} 
& \multicolumn{1}{c}{\rotatebox[origin=c]{270}{\textbf{\dsBinOne}}} 
& \multicolumn{1}{c}{\rotatebox[origin=c]{270}{\textbf{\dsBinTwo}}} 
& \multicolumn{1}{c}{\rotatebox[origin=c]{270}{\textbf{\dsBinThree}}} 
& \multicolumn{1}{c}{\rotatebox[origin=c]{270}{\textbf{\deltaPa}}} 
& \multicolumn{1}{c}{\rotatebox[origin=c]{270}{\textbf{\deltaPe}}} 
& \multicolumn{1}{c}{\rotatebox[origin=c]{270}{\textbf{\fL}}} 
& \multicolumn{1}{c}{\rotatebox[origin=c]{270}{\textbf{\fpa}}}\\
\midrule

\multicolumn{1}{@{}l}{\textbf{\ACPS}}       & $\phantom{-}1.00$ & $-0.09$           & $-0.07$           & $-0.27$           & $-0.18$           & $\phantom{-}0.04$ & $-0.01$           & $-0.13$           &$\phantom{-}0.15$ & $\phantom{-}0.05$ & $\phantom{-}0.02$ & $\phantom{-}0.07$ & $-0.01$           & $-0.00$           & $-0.01$           & $-0.04$ \\
\multicolumn{1}{@{}l}{\textbf{\ACPL}}       & $-0.09$           & $\phantom{-}1.00$ & $-0.16$           & $-0.08$           & $\phantom{-}0.01$ & $-0.02$           & $-0.03$           & $-0.02$           &$-0.01$           & $-0.03$           & $-0.03$           & $-0.02$           & $\phantom{-}0.01$ & $\phantom{-}0.02$ & $\phantom{-}0.01$ & $-0.01$ \\
\multicolumn{1}{@{}l}{\textbf{\ACPpa}}      & $-0.07$           & $-0.16$           & $\phantom{-}1.00$ & $-0.57$           & $-0.01$           & $-0.04$           & $-0.01$           & $-0.02$           &$-0.01$           & $-0.05$           & $-0.00$           & $\phantom{-}0.01$ & $-0.03$           & $-0.04$           & $-0.01$           & $\phantom{-}0.01$ \\
\multicolumn{1}{@{}l}{\textbf{\ACPpe}}      & $-0.27$           & $-0.08$           & $-0.57$           & $\phantom{-}1.00$ & $\phantom{-}0.06$ & $-0.01$           & $-0.03$           & $\phantom{-}0.04$ &$-0.04$           & $-0.00$           & $-0.02$           & $-0.02$           & $\phantom{-}0.01$ & $\phantom{-}0.02$ & $\phantom{-}0.01$ & $\phantom{-}0.04$ \\
\multicolumn{1}{@{}l}{\textbf{\AsBinZero}}  & $-0.18$           & $\phantom{-}0.01$ & $-0.01$           & $\phantom{-}0.06$ & $\phantom{-}1.00$ & $\phantom{-}0.00$ & $\phantom{-}0.02$ & $\phantom{-}0.04$ &$-0.65$           & $\phantom{-}0.00$ & $\phantom{-}0.01$ & $-0.02$           & $-0.00$           & $-0.03$           & $-0.00$           & $\phantom{-}0.14$ \\
\multicolumn{1}{@{}l}{\textbf{\AsBinOne}}   & $\phantom{-}0.04$ & $-0.02$           & $-0.04$           & $-0.01$           & $\phantom{-}0.00$ & $\phantom{-}1.00$ & $\phantom{-}0.03$ & $-0.00$           &$\phantom{-}0.00$ & $\phantom{-}0.73$ & $-0.01$           & $-0.01$           & $\phantom{-}0.10$ & $\phantom{-}0.09$ & $\phantom{-}0.13$ & $\phantom{-}0.01$ \\
\multicolumn{1}{@{}l}{\textbf{\AsBinTwo}}   & $-0.01$           & $-0.03$           & $-0.01$           & $-0.03$           & $\phantom{-}0.02$ & $\phantom{-}0.03$ & $\phantom{-}1.00$ & $\phantom{-}0.01$ &$-0.02$           & $\phantom{-}0.02$ & $\phantom{-}0.29$ & $-0.01$           & $\phantom{-}0.02$ & $\phantom{-}0.02$ & $\phantom{-}0.06$ & $\phantom{-}0.10$ \\
\multicolumn{1}{@{}l}{\textbf{\AsBinThree}} & $-0.13$           & $-0.02$           & $-0.02$           & $\phantom{-}0.04$ & $\phantom{-}0.04$ & $-0.00$           & $\phantom{-}0.01$ & $\phantom{-}1.00$ &$-0.03$           & $\phantom{-}0.01$ & $\phantom{-}0.01$ & $-0.22$           & $\phantom{-}0.02$ & $-0.02$           & $-0.04$           & $\phantom{-}0.10$ \\
\multicolumn{1}{@{}l}{\textbf{\dsBinZero}}  & $\phantom{-}0.15$ & $-0.01$           & $-0.01$           & $-0.04$           & $-0.65$           & $\phantom{-}0.00$ & $-0.02$           & $-0.03$           &$\phantom{-}1.00$ & $\phantom{-}0.01$ & $\phantom{-}0.00$ & $\phantom{-}0.02$ & $\phantom{-}0.06$ & $\phantom{-}0.06$ & $\phantom{-}0.01$ & $-0.14$ \\
\multicolumn{1}{@{}l}{\textbf{\dsBinOne}}   & $\phantom{-}0.05$ & $-0.03$           & $-0.05$           & $-0.00$           & $\phantom{-}0.00$ & $\phantom{-}0.73$ & $\phantom{-}0.02$ & $\phantom{-}0.01$ &$\phantom{-}0.01$ & $\phantom{-}1.00$ & $\phantom{-}0.04$ & $\phantom{-}0.02$ & $\phantom{-}0.17$ & $\phantom{-}0.14$ & $-0.04$           & $\phantom{-}0.08$ \\
\multicolumn{1}{@{}l}{\textbf{\dsBinTwo}}   & $\phantom{-}0.02$ & $-0.03$           & $-0.00$           & $-0.02$           & $\phantom{-}0.01$ & $-0.01$           & $\phantom{-}0.29$ & $\phantom{-}0.01$ &$\phantom{-}0.00$ & $\phantom{-}0.04$ & $\phantom{-}1.00$ & $\phantom{-}0.05$ & $\phantom{-}0.15$ & $\phantom{-}0.12$ & $-0.22$           & $\phantom{-}0.14$ \\
\multicolumn{1}{@{}l}{\textbf{\dsBinThree}} & $\phantom{-}0.07$ & $-0.02$           & $\phantom{-}0.01$ & $-0.02$           & $-0.02$           & $-0.01$           & $-0.01$           & $-0.22$           &$\phantom{-}0.02$ & $\phantom{-}0.02$ & $\phantom{-}0.05$ & $\phantom{-}1.00$ & $\phantom{-}0.12$ & $\phantom{-}0.11$ & $-0.14$           & $\phantom{-}0.02$ \\
\multicolumn{1}{@{}l}{\textbf{\deltaPa}}    & $-0.01$           & $\phantom{-}0.01$ & $-0.03$           & $\phantom{-}0.01$ & $-0.00$           & $\phantom{-}0.10$ & $\phantom{-}0.02$ & $\phantom{-}0.02$ &$\phantom{-}0.06$ & $\phantom{-}0.17$ & $\phantom{-}0.15$ & $\phantom{-}0.12$ & $\phantom{-}1.00$ & $\phantom{-}0.69$ & $-0.09$           & $\phantom{-}0.05$ \\
\multicolumn{1}{@{}l}{\textbf{\deltaPe}}    & $-0.00$           & $\phantom{-}0.02$ & $-0.04$           & $\phantom{-}0.02$ & $-0.03$           & $\phantom{-}0.09$ & $\phantom{-}0.02$ & $-0.02$           &$\phantom{-}0.06$ & $\phantom{-}0.14$ & $\phantom{-}0.12$ & $\phantom{-}0.11$ & $\phantom{-}0.69$ & $\phantom{-}1.00$ & $-0.03$           & $\phantom{-}0.00$ \\
\multicolumn{1}{@{}l}{\textbf{\fL}}         & $-0.01$           & $\phantom{-}0.01$ & $-0.01$           & $\phantom{-}0.01$ & $-0.00$           & $\phantom{-}0.13$ & $\phantom{-}0.06$ & $-0.04$           &$\phantom{-}0.01$ & $-0.04$           & $-0.22$           & $-0.14$           & $-0.09$           & $-0.03$           & $\phantom{-}1.00$ & $-0.44$ \\
\multicolumn{1}{@{}l}{\textbf{\fpa}}        & $-0.04$           & $-0.01$           & $\phantom{-}0.01$ & $\phantom{-}0.04$ & $\phantom{-}0.14$ & $\phantom{-}0.01$ & $\phantom{-}0.10$ & $\phantom{-}0.10$ &$-0.14$           & $\phantom{-}0.08$ & $\phantom{-}0.14$ & $\phantom{-}0.02$ & $\phantom{-}0.05$ & $\phantom{-}0.00$ & $-0.44$           & $\phantom{-}1.00$ \\

\bottomrule

\end{tabularx}

} % <-- Close the \rotatebox

\label{tab:angAna:data:Bs:correlationMatrix:SPwave}

\end{table}

% This should be taken out in the final paper
% \input{supplementary-app}

\addcontentsline{toc}{section}{References}
%\setboolean{inbibliography}{true}
\bibliographystyle{LHCb}
\bibliography{main,standard,LHCb-PAPER,LHCb-CONF,LHCb-DP,LHCb-TDR}

\ifx\mcitethebibliography\mciteundefinedmacro
\PackageError{LHCb.bst}{mciteplus.sty has not been loaded}
{This bibstyle requires the use of the mciteplus package.}\fi
\providecommand{\href}[2]{#2}
\begin{mcitethebibliography}{10}
\mciteSetBstSublistMode{n}
\mciteSetBstMaxWidthForm{subitem}{\alph{mcitesubitemcount})}
\mciteSetBstSublistLabelBeginEnd{\mcitemaxwidthsubitemform\space}
{\relax}{\relax}

\bibitem{LHCb-PAPER-2011-021}
LHCb collaboration, R.~Aaij {\em et~al.}, \ifthenelse{\boolean{articletitles}}{\emph{{Measurement of the \CP-violating phase \phis in the decay \mbox{\decay{\Bs}{\jpsi\phiz}}}}, }{}\href{https://doi.org/10.1103/PhysRevLett.108.101803}{Phys.\ Rev.\ Lett.\  \textbf{108} (2012) 101803}, \href{http://arxiv.org/abs/1112.3183}{{\normalfont\ttfamily arXiv:1112.3183}}\relax
\mciteBstWouldAddEndPuncttrue
\mciteSetBstMidEndSepPunct{\mcitedefaultmidpunct}
{\mcitedefaultendpunct}{\mcitedefaultseppunct}\relax
\EndOfBibitem
\bibitem{DeBruyn:2022zhw}
K.~De~Bruyn, R.~Fleischer, E.~Malami, and P.~van Vliet, \ifthenelse{\boolean{articletitles}}{\emph{{New physics in $B_{q}^{0}$\textendash{}$\overline{B}_{q}^{0}$ mixing: present challenges, prospects, and implications for}}, }{}\href{https://doi.org/10.1088/1361-6471/acab1d}{J.\ Phys.\  \textbf{G50} (2023) 045003}, \href{http://arxiv.org/abs/2208.14910}{{\normalfont\ttfamily arXiv:2208.14910}}\relax
\mciteBstWouldAddEndPuncttrue
\mciteSetBstMidEndSepPunct{\mcitedefaultmidpunct}
{\mcitedefaultendpunct}{\mcitedefaultseppunct}\relax
\EndOfBibitem
\bibitem{BIGI198185}
I.~I. Bigi and A.~I. Sanada, \ifthenelse{\boolean{articletitles}}{\emph{{Notes on the observability of CP violations in B decays}}, }{}\href{https://doi.org/https://doi.org/10.1016/0550-3213(81)90519-8}{Nucl.\ Phys.\  \textbf{B193} (1981) 85}\relax
\mciteBstWouldAddEndPuncttrue
\mciteSetBstMidEndSepPunct{\mcitedefaultmidpunct}
{\mcitedefaultendpunct}{\mcitedefaultseppunct}\relax
\EndOfBibitem
\bibitem{PhysRevD.23.1567}
A.~B. Carter and A.~I. Sanda, \ifthenelse{\boolean{articletitles}}{\emph{{$\CP$ violation in $B$-meson decays}}, }{}\href{https://doi.org/10.1103/PhysRevD.23.1567}{Phys.\ Rev.\  \textbf{{D23}} (1981) 1567}\relax
\mciteBstWouldAddEndPuncttrue
\mciteSetBstMidEndSepPunct{\mcitedefaultmidpunct}
{\mcitedefaultendpunct}{\mcitedefaultseppunct}\relax
\EndOfBibitem
\bibitem{PhysRevLett.45.952}
A.~B. Carter and A.~I. Sanda, \ifthenelse{\boolean{articletitles}}{\emph{{$\CP$ nonconservation in cascade decays of $B$ mesons}}, }{}\href{https://doi.org/10.1103/PhysRevLett.45.952}{Phys.\ Rev.\ Lett.\  \textbf{45} (1980) 952}\relax
\mciteBstWouldAddEndPuncttrue
\mciteSetBstMidEndSepPunct{\mcitedefaultmidpunct}
{\mcitedefaultendpunct}{\mcitedefaultseppunct}\relax
\EndOfBibitem
\bibitem{CKMfitter2015}
CKMfitter group, J.~Charles {\em et~al.}, \ifthenelse{\boolean{articletitles}}{\emph{{Current status of the standard model CKM fit and constraints on \hbox{$\Delta F=2$} new physics}}, }{}\href{https://doi.org/10.1103/PhysRevD.91.073007}{Phys.\ Rev.\  \textbf{D91} (2015) 073007}, \href{http://arxiv.org/abs/1501.05013}{{\normalfont\ttfamily arXiv:1501.05013}}, {updated results and plots available at \href{http://ckmfitter.in2p3.fr/}{{\texttt{http://ckmfitter.in2p3.fr/}}}}\relax
\mciteBstWouldAddEndPuncttrue
\mciteSetBstMidEndSepPunct{\mcitedefaultmidpunct}
{\mcitedefaultendpunct}{\mcitedefaultseppunct}\relax
\EndOfBibitem
\bibitem{UTfit-UT}
UTfit collaboration, M.~Bona {\em et~al.}, \ifthenelse{\boolean{articletitles}}{\emph{{The unitarity triangle fit in the standard model and hadronic parameters from lattice QCD: A reappraisal after the measurements of $\Delta m_{s}$ and $BR(B\to\tau\nu_{\tau})$}}, }{}\href{https://doi.org/10.1088/1126-6708/2006/10/081}{JHEP \textbf{10} (2006) 081}, \href{http://arxiv.org/abs/hep-ph/0606167}{{\normalfont\ttfamily arXiv:hep-ph/0606167}}, {updated results and plots available at \href{http://www.utfit.org/}{{\texttt{http://www.utfit.org/}}}}\relax
\mciteBstWouldAddEndPuncttrue
\mciteSetBstMidEndSepPunct{\mcitedefaultmidpunct}
{\mcitedefaultendpunct}{\mcitedefaultseppunct}\relax
\EndOfBibitem
\bibitem{HFLAV23}
Heavy Flavor Averaging Group, S.~Banerjee {\em et~al.}, \ifthenelse{\boolean{articletitles}}{\emph{{Averages of $b$-hadron, $c$-hadron, and $\tau$-lepton properties as of 2023}}, }{}\href{http://arxiv.org/abs/2411.18639}{{\normalfont\ttfamily arXiv:2411.18639}}, {updated results and plots available at \href{https://hflav.web.cern.ch}{{\texttt{https://hflav.web.cern.ch}}}}\relax
\mciteBstWouldAddEndPuncttrue
\mciteSetBstMidEndSepPunct{\mcitedefaultmidpunct}
{\mcitedefaultendpunct}{\mcitedefaultseppunct}\relax
\EndOfBibitem
\bibitem{Chiang2009NewPI}
C.-W. Chiang {\em et~al.}, \ifthenelse{\boolean{articletitles}}{\emph{{New physics in $B^{0}_{s} \to J/\psi \phi$: a general analysis}}, }{}\href{https://doi.org/10.1007/JHEP04(2010)031}{JHEP \textbf{04} (2010) 031}, \href{http://arxiv.org/abs/0910.2929}{{\normalfont\ttfamily arXiv:0910.2929}}\relax
\mciteBstWouldAddEndPuncttrue
\mciteSetBstMidEndSepPunct{\mcitedefaultmidpunct}
{\mcitedefaultendpunct}{\mcitedefaultseppunct}\relax
\EndOfBibitem
\bibitem{Liu:2013nea}
X.~Liu, W.~Wang, and Y.~Xie, \ifthenelse{\boolean{articletitles}}{\emph{{Penguin pollution in $B\to J/\psi V$ decays and impact on the extraction of the $\Bs$--$\Bsb$ mixing phase}}, }{}\href{https://doi.org/10.1103/PhysRevD.89.094010}{Phys.\ Rev.\  \textbf{D89} (2014) 094010}, \href{http://arxiv.org/abs/1309.0313}{{\normalfont\ttfamily arXiv:1309.0313}}\relax
\mciteBstWouldAddEndPuncttrue
\mciteSetBstMidEndSepPunct{\mcitedefaultmidpunct}
{\mcitedefaultendpunct}{\mcitedefaultseppunct}\relax
\EndOfBibitem
\bibitem{Frings:2015eva}
P.~Frings, U.~Nierste, and M.~Wiebusch, \ifthenelse{\boolean{articletitles}}{\emph{{Penguin contributions to CP phases in $B_{d,s}$ decays to charmonium}}, }{}\href{https://doi.org/10.1103/PhysRevLett.115.061802}{Phys.\ Rev.\ Lett.\  \textbf{115} (2015) 061802}, \href{http://arxiv.org/abs/1503.00859}{{\normalfont\ttfamily arXiv:1503.00859}}\relax
\mciteBstWouldAddEndPuncttrue
\mciteSetBstMidEndSepPunct{\mcitedefaultmidpunct}
{\mcitedefaultendpunct}{\mcitedefaultseppunct}\relax
\EndOfBibitem
\bibitem{Fleischer:1999zi}
R.~Fleischer, \ifthenelse{\boolean{articletitles}}{\emph{{Extracting CKM phases from angular distributions of $B_{d,s}$ decays into admixtures of CP eigenstates}}, }{}\href{https://doi.org/10.1103/PhysRevD.60.073008}{Phys.\ Rev.\  \textbf{D60} (1999) 073008}, \href{http://arxiv.org/abs/hep-ph/9903540}{{\normalfont\ttfamily arXiv:hep-ph/9903540}}\relax
\mciteBstWouldAddEndPuncttrue
\mciteSetBstMidEndSepPunct{\mcitedefaultmidpunct}
{\mcitedefaultendpunct}{\mcitedefaultseppunct}\relax
\EndOfBibitem
\bibitem{Faller:2008gt}
S.~Faller, R.~Fleischer, and T.~Mannel, \ifthenelse{\boolean{articletitles}}{\emph{{Precision physics with $B^0_s \to J/\psi \phi$ at the LHC: the quest for new physics}}, }{}\href{https://doi.org/10.1103/PhysRevD.79.014005}{Phys.\ Rev.\  \textbf{D79} (2009) 014005}, \href{http://arxiv.org/abs/0810.4248}{{\normalfont\ttfamily arXiv:0810.4248}}\relax
\mciteBstWouldAddEndPuncttrue
\mciteSetBstMidEndSepPunct{\mcitedefaultmidpunct}
{\mcitedefaultendpunct}{\mcitedefaultseppunct}\relax
\EndOfBibitem
\bibitem{DeBruyn:2014oga}
K.~De~Bruyn and R.~Fleischer, \ifthenelse{\boolean{articletitles}}{\emph{{A roadmap to control penguin effects in $B^0_d\to J/\psi K_{\rm S}^0$ and $B^0_s\to J/\psi \phi$}}, }{}\href{https://doi.org/10.1007/JHEP03(2015)145}{JHEP \textbf{03} (2015) 145}, \href{http://arxiv.org/abs/1412.6834}{{\normalfont\ttfamily arXiv:1412.6834}}\relax
\mciteBstWouldAddEndPuncttrue
\mciteSetBstMidEndSepPunct{\mcitedefaultmidpunct}
{\mcitedefaultendpunct}{\mcitedefaultseppunct}\relax
\EndOfBibitem
\bibitem{Barel:2020jvf}
M.~Z. Barel, K.~De~Bruyn, R.~Fleischer, and E.~Malami, \ifthenelse{\boolean{articletitles}}{\emph{{In pursuit of new physics with $B_d^0\to J/\psi K^0$ and $B_s^0\to J/\psi\phi$ decays at the high-precision frontier}}, }{}\href{https://doi.org/10.1088/1361-6471/abf2a2}{J.\ Phys.\  \textbf{G48} (2021) 065002}, \href{http://arxiv.org/abs/2010.14423}{{\normalfont\ttfamily arXiv:2010.14423}}\relax
\mciteBstWouldAddEndPuncttrue
\mciteSetBstMidEndSepPunct{\mcitedefaultmidpunct}
{\mcitedefaultendpunct}{\mcitedefaultseppunct}\relax
\EndOfBibitem
\bibitem{DeBruyn:2025rhk}
K.~De~Bruyn, R.~Fleischer, and E.~Malami, \ifthenelse{\boolean{articletitles}}{\emph{{How to tame penguins: Advancing to high-precision measurements of $\phi_d$ and $\phi_s$}}, }{}\href{http://arxiv.org/abs/2505.06102}{{\normalfont\ttfamily arXiv:2505.06102}}\relax
\mciteBstWouldAddEndPuncttrue
\mciteSetBstMidEndSepPunct{\mcitedefaultmidpunct}
{\mcitedefaultendpunct}{\mcitedefaultseppunct}\relax
\EndOfBibitem
\bibitem{LHCb-PAPER-2015-034}
LHCb collaboration, R.~Aaij {\em et~al.}, \ifthenelse{\boolean{articletitles}}{\emph{{Measurement of \CP violation parameters and polarisation fractions in \mbox{\decay{\Bs}{\jpsi\Kstarzb}} decays}}, }{}\href{https://doi.org/10.1007/JHEP11(2015)082}{JHEP \textbf{11} (2015) 082}, \href{http://arxiv.org/abs/1509.00400}{{\normalfont\ttfamily arXiv:1509.00400}}\relax
\mciteBstWouldAddEndPuncttrue
\mciteSetBstMidEndSepPunct{\mcitedefaultmidpunct}
{\mcitedefaultendpunct}{\mcitedefaultseppunct}\relax
\EndOfBibitem
\bibitem{LHCb-PAPER-2014-058}
LHCb collaboration, R.~Aaij {\em et~al.}, \ifthenelse{\boolean{articletitles}}{\emph{{Measurement of the \CP-violating phase $\beta$ in \mbox{\decay{\Bzb}{\jpsi\pip\pim}} decays and limits on penguin effects}}, }{}\href{https://doi.org/10.1016/j.physletb.2015.01.008}{Phys.\ Lett.\  \textbf{B742} (2015) 38}, \href{http://arxiv.org/abs/1411.1634}{{\normalfont\ttfamily arXiv:1411.1634}}\relax
\mciteBstWouldAddEndPuncttrue
\mciteSetBstMidEndSepPunct{\mcitedefaultmidpunct}
{\mcitedefaultendpunct}{\mcitedefaultseppunct}\relax
\EndOfBibitem
\bibitem{LHCb-DP-2008-001}
LHCb collaboration, A.~A. Alves~Jr.\ {\em et~al.}, \ifthenelse{\boolean{articletitles}}{\emph{{The \lhcb detector at the LHC}}, }{}\href{https://doi.org/10.1088/1748-0221/3/08/S08005}{JINST \textbf{3} (2008) S08005}\relax
\mciteBstWouldAddEndPuncttrue
\mciteSetBstMidEndSepPunct{\mcitedefaultmidpunct}
{\mcitedefaultendpunct}{\mcitedefaultseppunct}\relax
\EndOfBibitem
\bibitem{LHCb-DP-2014-002}
LHCb collaboration, R.~Aaij {\em et~al.}, \ifthenelse{\boolean{articletitles}}{\emph{{LHCb detector performance}}, }{}\href{https://doi.org/10.1142/S0217751X15300227}{Int.\ J.\ Mod.\ Phys.\  \textbf{A30} (2015) 1530022}, \href{http://arxiv.org/abs/1412.6352}{{\normalfont\ttfamily arXiv:1412.6352}}\relax
\mciteBstWouldAddEndPuncttrue
\mciteSetBstMidEndSepPunct{\mcitedefaultmidpunct}
{\mcitedefaultendpunct}{\mcitedefaultseppunct}\relax
\EndOfBibitem
\bibitem{LHCb-DP-2014-001}
R.~Aaij {\em et~al.}, \ifthenelse{\boolean{articletitles}}{\emph{{Performance of the LHCb Vertex Locator}}, }{}\href{https://doi.org/10.1088/1748-0221/9/09/P09007}{JINST \textbf{9} (2014) P09007}, \href{http://arxiv.org/abs/1405.7808}{{\normalfont\ttfamily arXiv:1405.7808}}\relax
\mciteBstWouldAddEndPuncttrue
\mciteSetBstMidEndSepPunct{\mcitedefaultmidpunct}
{\mcitedefaultendpunct}{\mcitedefaultseppunct}\relax
\EndOfBibitem
\bibitem{LHCb-DP-2017-001}
P.~d'Argent {\em et~al.}, \ifthenelse{\boolean{articletitles}}{\emph{{Improved performance of the LHCb Outer Tracker in LHC Run 2}}, }{}\href{https://doi.org/10.1088/1748-0221/12/11/P11016}{JINST \textbf{12} (2017) P11016}, \href{http://arxiv.org/abs/1708.00819}{{\normalfont\ttfamily arXiv:1708.00819}}\relax
\mciteBstWouldAddEndPuncttrue
\mciteSetBstMidEndSepPunct{\mcitedefaultmidpunct}
{\mcitedefaultendpunct}{\mcitedefaultseppunct}\relax
\EndOfBibitem
\bibitem{LHCb-DP-2012-003}
M.~Adinolfi {\em et~al.}, \ifthenelse{\boolean{articletitles}}{\emph{{Performance of the \lhcb RICH detector at the LHC}}, }{}\href{https://doi.org/10.1140/epjc/s10052-013-2431-9}{Eur.\ Phys.\ J.\  \textbf{C73} (2013) 2431}, \href{http://arxiv.org/abs/1211.6759}{{\normalfont\ttfamily arXiv:1211.6759}}\relax
\mciteBstWouldAddEndPuncttrue
\mciteSetBstMidEndSepPunct{\mcitedefaultmidpunct}
{\mcitedefaultendpunct}{\mcitedefaultseppunct}\relax
\EndOfBibitem
\bibitem{LHCb-DP-2012-002}
A.~A. Alves~Jr.\ {\em et~al.}, \ifthenelse{\boolean{articletitles}}{\emph{{Performance of the LHCb muon system}}, }{}\href{https://doi.org/10.1088/1748-0221/8/02/P02022}{JINST \textbf{8} (2013) P02022}, \href{http://arxiv.org/abs/1211.1346}{{\normalfont\ttfamily arXiv:1211.1346}}\relax
\mciteBstWouldAddEndPuncttrue
\mciteSetBstMidEndSepPunct{\mcitedefaultmidpunct}
{\mcitedefaultendpunct}{\mcitedefaultseppunct}\relax
\EndOfBibitem
\bibitem{Sjostrand:2007gs}
T.~Sj\"{o}strand, S.~Mrenna, and P.~Skands, \ifthenelse{\boolean{articletitles}}{\emph{{A brief introduction to PYTHIA 8.1}}, }{}\href{https://doi.org/10.1016/j.cpc.2008.01.036}{Comput.\ Phys.\ Commun.\  \textbf{178} (2008) 852}, \href{http://arxiv.org/abs/0710.3820}{{\normalfont\ttfamily arXiv:0710.3820}}\relax
\mciteBstWouldAddEndPuncttrue
\mciteSetBstMidEndSepPunct{\mcitedefaultmidpunct}
{\mcitedefaultendpunct}{\mcitedefaultseppunct}\relax
\EndOfBibitem
\bibitem{Sjostrand:2006za}
T.~Sj\"{o}strand, S.~Mrenna, and P.~Skands, \ifthenelse{\boolean{articletitles}}{\emph{{PYTHIA 6.4 physics and manual}}, }{}\href{https://doi.org/10.1088/1126-6708/2006/05/026}{JHEP \textbf{05} (2006) 026}, \href{http://arxiv.org/abs/hep-ph/0603175}{{\normalfont\ttfamily arXiv:hep-ph/0603175}}\relax
\mciteBstWouldAddEndPuncttrue
\mciteSetBstMidEndSepPunct{\mcitedefaultmidpunct}
{\mcitedefaultendpunct}{\mcitedefaultseppunct}\relax
\EndOfBibitem
\bibitem{LHCb-PROC-2010-056}
I.~Belyaev {\em et~al.}, \ifthenelse{\boolean{articletitles}}{\emph{{Handling of the generation of primary events in Gauss, the LHCb simulation framework}}, }{}\href{https://doi.org/10.1088/1742-6596/331/3/032047}{J.\ Phys.\ Conf.\ Ser.\  \textbf{331} (2011) 032047}\relax
\mciteBstWouldAddEndPuncttrue
\mciteSetBstMidEndSepPunct{\mcitedefaultmidpunct}
{\mcitedefaultendpunct}{\mcitedefaultseppunct}\relax
\EndOfBibitem
\bibitem{Lange:2001uf}
D.~J. Lange, \ifthenelse{\boolean{articletitles}}{\emph{{The EvtGen particle decay simulation package}}, }{}\href{https://doi.org/10.1016/S0168-9002(01)00089-4}{Nucl.\ Instrum.\ Meth.\  \textbf{A462} (2001) 152}\relax
\mciteBstWouldAddEndPuncttrue
\mciteSetBstMidEndSepPunct{\mcitedefaultmidpunct}
{\mcitedefaultendpunct}{\mcitedefaultseppunct}\relax
\EndOfBibitem
\bibitem{davidson2015photos}
N.~Davidson, T.~Przedzinski, and Z.~Was, \ifthenelse{\boolean{articletitles}}{\emph{{PHOTOS interface in C++: Technical and physics documentation}}, }{}\href{https://doi.org/https://doi.org/10.1016/j.cpc.2015.09.013}{Comp.\ Phys.\ Comm.\  \textbf{199} (2016) 86}, \href{http://arxiv.org/abs/1011.0937}{{\normalfont\ttfamily arXiv:1011.0937}}\relax
\mciteBstWouldAddEndPuncttrue
\mciteSetBstMidEndSepPunct{\mcitedefaultmidpunct}
{\mcitedefaultendpunct}{\mcitedefaultseppunct}\relax
\EndOfBibitem
\bibitem{Allison:2006ve}
Geant4 collaboration, J.~Allison {\em et~al.}, \ifthenelse{\boolean{articletitles}}{\emph{{Geant4 developments and applications}}, }{}\href{https://doi.org/10.1109/TNS.2006.869826}{IEEE Trans.\ Nucl.\ Sci.\  \textbf{53} (2006) 270}\relax
\mciteBstWouldAddEndPuncttrue
\mciteSetBstMidEndSepPunct{\mcitedefaultmidpunct}
{\mcitedefaultendpunct}{\mcitedefaultseppunct}\relax
\EndOfBibitem
\bibitem{Agostinelli:2002hh}
Geant4 collaboration, S.~Agostinelli {\em et~al.}, \ifthenelse{\boolean{articletitles}}{\emph{{Geant4: A simulation toolkit}}, }{}\href{https://doi.org/10.1016/S0168-9002(03)01368-8}{Nucl.\ Instrum.\ Meth.\  \textbf{A506} (2003) 250}\relax
\mciteBstWouldAddEndPuncttrue
\mciteSetBstMidEndSepPunct{\mcitedefaultmidpunct}
{\mcitedefaultendpunct}{\mcitedefaultseppunct}\relax
\EndOfBibitem
\bibitem{LHCb-PROC-2011-006}
M.~Clemencic {\em et~al.}, \ifthenelse{\boolean{articletitles}}{\emph{{The \lhcb simulation application, Gauss: Design, evolution and experience}}, }{}\href{https://doi.org/10.1088/1742-6596/331/3/032023}{J.\ Phys.\ Conf.\ Ser.\  \textbf{331} (2011) 032023}\relax
\mciteBstWouldAddEndPuncttrue
\mciteSetBstMidEndSepPunct{\mcitedefaultmidpunct}
{\mcitedefaultendpunct}{\mcitedefaultseppunct}\relax
\EndOfBibitem
\bibitem{Rogozhnikov:2016bdp}
A.~Rogozhnikov, \ifthenelse{\boolean{articletitles}}{\emph{{Reweighting with boosted decision trees}}, }{}\href{https://doi.org/10.1088/1742-6596/762/1/012036}{J.\ Phys.\ Conf.\ Ser.\  \textbf{762} (2016) 012036}, \href{http://arxiv.org/abs/1608.05806}{{\normalfont\ttfamily arXiv:1608.05806}}, \url{https://github.com/arogozhnikov/hep_ml}\relax
\mciteBstWouldAddEndPuncttrue
\mciteSetBstMidEndSepPunct{\mcitedefaultmidpunct}
{\mcitedefaultendpunct}{\mcitedefaultseppunct}\relax
\EndOfBibitem
\bibitem{LHCb-PUB-2016-021}
L.~Anderlini {\em et~al.}, \ifthenelse{\boolean{articletitles}}{\emph{{The PIDCalib package}}, }{} \href{http://cdsweb.cern.ch/search?p=LHCb-PUB-2016-021&f=reportnumber&action_search=Search&c=LHCb+Notes} {LHCb-PUB-2016-021}, 2016\relax
\mciteBstWouldAddEndPuncttrue
\mciteSetBstMidEndSepPunct{\mcitedefaultmidpunct}
{\mcitedefaultendpunct}{\mcitedefaultseppunct}\relax
\EndOfBibitem
\bibitem{LHCb-DP-2018-001}
R.~Aaij {\em et~al.}, \ifthenelse{\boolean{articletitles}}{\emph{{Selection and processing of calibration samples to measure the particle identification performance of the LHCb experiment in Run 2}}, }{}\href{https://doi.org/10.1140/epjti/s40485-019-0050-z}{Eur.\ Phys.\ J.\ Tech.\ Instr.\  \textbf{6} (2019) 1}, \href{http://arxiv.org/abs/1803.00824}{{\normalfont\ttfamily arXiv:1803.00824}}\relax
\mciteBstWouldAddEndPuncttrue
\mciteSetBstMidEndSepPunct{\mcitedefaultmidpunct}
{\mcitedefaultendpunct}{\mcitedefaultseppunct}\relax
\EndOfBibitem
\bibitem{LHCb-DP-2012-004}
R.~Aaij {\em et~al.}, \ifthenelse{\boolean{articletitles}}{\emph{{The \lhcb trigger and its performance in 2011}}, }{}\href{https://doi.org/10.1088/1748-0221/8/04/P04022}{JINST \textbf{8} (2013) P04022}, \href{http://arxiv.org/abs/1211.3055}{{\normalfont\ttfamily arXiv:1211.3055}}\relax
\mciteBstWouldAddEndPuncttrue
\mciteSetBstMidEndSepPunct{\mcitedefaultmidpunct}
{\mcitedefaultendpunct}{\mcitedefaultseppunct}\relax
\EndOfBibitem
\bibitem{LHCb-PAPER-2019-042}
LHCb collaboration, R.~Aaij {\em et~al.}, \ifthenelse{\boolean{articletitles}}{\emph{{First observation of excited $\Omegares_b^-$ states}}, }{}\href{https://doi.org/10.1103/PhysRevLett.124.082002}{Phys.\ Rev.\ Lett.\  \textbf{124} (2020) 082002}, \href{http://arxiv.org/abs/2001.00851}{{\normalfont\ttfamily arXiv:2001.00851}}\relax
\mciteBstWouldAddEndPuncttrue
\mciteSetBstMidEndSepPunct{\mcitedefaultmidpunct}
{\mcitedefaultendpunct}{\mcitedefaultseppunct}\relax
\EndOfBibitem
\bibitem{PDG2024}
Particle Data Group, S.~Navas {\em et~al.}, \ifthenelse{\boolean{articletitles}}{\emph{{\href{http://pdg.lbl.gov/}{Review of particle physics}}}, }{}\href{https://doi.org/10.1103/PhysRevD.110.030001}{Phys.\ Rev.\  \textbf{D110} (2024) 030001}\relax
\mciteBstWouldAddEndPuncttrue
\mciteSetBstMidEndSepPunct{\mcitedefaultmidpunct}
{\mcitedefaultendpunct}{\mcitedefaultseppunct}\relax
\EndOfBibitem
\bibitem{Hulsbergen:2005pu}
W.~D. Hulsbergen, \ifthenelse{\boolean{articletitles}}{\emph{{Decay chain fitting with a Kalman filter}}, }{}\href{https://doi.org/10.1016/j.nima.2005.06.078}{Nucl.\ Instrum.\ Meth.\  \textbf{A552} (2005) 566}, \href{http://arxiv.org/abs/physics/0503191}{{\normalfont\ttfamily arXiv:physics/0503191}}\relax
\mciteBstWouldAddEndPuncttrue
\mciteSetBstMidEndSepPunct{\mcitedefaultmidpunct}
{\mcitedefaultendpunct}{\mcitedefaultseppunct}\relax
\EndOfBibitem
\bibitem{Breiman}
L.~Breiman, J.~H. Friedman, R.~A. Olshen, and C.~J. Stone, {\em Classification and regression trees}, Wadsworth international group, Belmont, California, USA, 1984\relax
\mciteBstWouldAddEndPuncttrue
\mciteSetBstMidEndSepPunct{\mcitedefaultmidpunct}
{\mcitedefaultendpunct}{\mcitedefaultseppunct}\relax
\EndOfBibitem
\bibitem{kFold_modified}
A.~Blum, A.~Kalai, and J.~Langford, \ifthenelse{\boolean{articletitles}}{\emph{Beating the hold-out: bounds for k-fold and progressive cross-validation}, }{} in {\em Proceedings of the Twelfth Annual Conference on Computational Learning Theory}, \href{https://doi.org/10.1145/307400.307439}{ COLT '99, {203208}, Association for Computing Machinery, 1999}\relax
\mciteBstWouldAddEndPuncttrue
\mciteSetBstMidEndSepPunct{\mcitedefaultmidpunct}
{\mcitedefaultendpunct}{\mcitedefaultseppunct}\relax
\EndOfBibitem
\bibitem{Koppenburg:2017zsh}
P.~Koppenburg, \ifthenelse{\boolean{articletitles}}{\emph{{Statistical biases in measurements with multiple candidates}}, }{}\href{http://arxiv.org/abs/1703.01128}{{\normalfont\ttfamily arXiv:1703.01128}}\relax
\mciteBstWouldAddEndPuncttrue
\mciteSetBstMidEndSepPunct{\mcitedefaultmidpunct}
{\mcitedefaultendpunct}{\mcitedefaultseppunct}\relax
\EndOfBibitem
\bibitem{LHCb-PAPER-2016-031}
LHCb collaboration, R.~Aaij {\em et~al.}, \ifthenelse{\boolean{articletitles}}{\emph{{Measurement of the \bquark-quark production cross-section in 7 and 13$~\tev$ $\proton\proton$ collisions}}, }{}\href{https://doi.org/10.1103/PhysRevLett.118.052002}{Phys.\ Rev.\ Lett.\  \textbf{118} (2017) 052002}, Erratum \href{https://doi.org/10.1103/PhysRevLett.119.169901}{ibid.\   \textbf{119} (2017) 169901}, \href{http://arxiv.org/abs/1612.05140}{{\normalfont\ttfamily arXiv:1612.05140}}\relax
\mciteBstWouldAddEndPuncttrue
\mciteSetBstMidEndSepPunct{\mcitedefaultmidpunct}
{\mcitedefaultendpunct}{\mcitedefaultseppunct}\relax
\EndOfBibitem
\bibitem{LHCb-PAPER-2020-046}
LHCb collaboration, R.~Aaij {\em et~al.}, \ifthenelse{\boolean{articletitles}}{\emph{{Precise measurement of the $f_s/f_d$ ratio of fragmentation fractions and of $B^0_s$ decay branching fractions}}, }{}\href{https://doi.org/10.1103/PhysRevD.104.032005}{Phys.\ Rev.\  \textbf{D104} (2021) 032005}, \href{http://arxiv.org/abs/2103.06810}{{\normalfont\ttfamily arXiv:2103.06810}}\relax
\mciteBstWouldAddEndPuncttrue
\mciteSetBstMidEndSepPunct{\mcitedefaultmidpunct}
{\mcitedefaultendpunct}{\mcitedefaultseppunct}\relax
\EndOfBibitem
\bibitem{LHCb-PAPER-2014-012}
LHCb collaboration, R.~Aaij {\em et~al.}, \ifthenelse{\boolean{articletitles}}{\emph{{Measurement of the resonant and \CP components in \mbox{\decay{\Bzb}{\jpsi\pip\pim}} decays}}, }{}\href{https://doi.org/10.1103/PhysRevD.90.012003}{Phys.\ Rev.\  \textbf{D90} (2014) 012003}, \href{http://arxiv.org/abs/1404.5673}{{\normalfont\ttfamily arXiv:1404.5673}}\relax
\mciteBstWouldAddEndPuncttrue
\mciteSetBstMidEndSepPunct{\mcitedefaultmidpunct}
{\mcitedefaultendpunct}{\mcitedefaultseppunct}\relax
\EndOfBibitem
\bibitem{LHCb-PAPER-2012-040}
LHCb collaboration, R.~Aaij {\em et~al.}, \ifthenelse{\boolean{articletitles}}{\emph{{Amplitude analysis and branching fraction measurement of \mbox{\decay{\Bsb}{\jpsi\Kp\Km}}}}, }{}\href{https://doi.org/10.1103/PhysRevD.87.072004}{Phys.\ Rev.\  \textbf{D87} (2013) 072004}, \href{http://arxiv.org/abs/1302.1213}{{\normalfont\ttfamily arXiv:1302.1213}}\relax
\mciteBstWouldAddEndPuncttrue
\mciteSetBstMidEndSepPunct{\mcitedefaultmidpunct}
{\mcitedefaultendpunct}{\mcitedefaultseppunct}\relax
\EndOfBibitem
\bibitem{LHCb-PAPER-2015-029}
LHCb collaboration, R.~Aaij {\em et~al.}, \ifthenelse{\boolean{articletitles}}{\emph{{Observation of $\jpsi\proton$ resonances consistent with pentaquark states in \mbox{\decay{\Lb}{\jpsi\proton\Km}} decays}}, }{}\href{https://doi.org/10.1103/PhysRevLett.115.072001}{Phys.\ Rev.\ Lett.\  \textbf{115} (2015) 072001}, \href{http://arxiv.org/abs/1507.03414}{{\normalfont\ttfamily arXiv:1507.03414}}\relax
\mciteBstWouldAddEndPuncttrue
\mciteSetBstMidEndSepPunct{\mcitedefaultmidpunct}
{\mcitedefaultendpunct}{\mcitedefaultseppunct}\relax
\EndOfBibitem
\bibitem{DeCian:2255039}
M.~De~Cian, S.~Farry, P.~Seyfert, and S.~Stahl, \ifthenelse{\boolean{articletitles}}{\emph{{Fast neural-net based fake track rejection in the LHCb reconstruction}}, }{} \href{http://cdsweb.cern.ch/search?p=LHCb-PUB-2017-011&f=reportnumber&action_search=Search&c=LHCb+Notes} {LHCb-PUB-2017-011}, 2017\relax
\mciteBstWouldAddEndPuncttrue
\mciteSetBstMidEndSepPunct{\mcitedefaultmidpunct}
{\mcitedefaultendpunct}{\mcitedefaultseppunct}\relax
\EndOfBibitem
\bibitem{Pivk:2004ty}
M.~Pivk and F.~R. Le~Diberder, \ifthenelse{\boolean{articletitles}}{\emph{{sPlot: A statistical tool to unfold data distributions}}, }{}\href{https://doi.org/10.1016/j.nima.2005.08.106}{Nucl.\ Instrum.\ Meth.\  \textbf{A555} (2005) 356}, \href{http://arxiv.org/abs/physics/0402083}{{\normalfont\ttfamily arXiv:physics/0402083}}\relax
\mciteBstWouldAddEndPuncttrue
\mciteSetBstMidEndSepPunct{\mcitedefaultmidpunct}
{\mcitedefaultendpunct}{\mcitedefaultseppunct}\relax
\EndOfBibitem
\bibitem{Skwarnicki:1986xj}
T.~Skwarnicki, {\em {A study of the radiative cascade transitions between the Upsilon-prime and Upsilon resonances}}, PhD thesis, Institute of Nuclear Physics, Krakow, 1986, {\href{http://inspirehep.net/record/230779/}{DESY-F31-86-02}}\relax
\mciteBstWouldAddEndPuncttrue
\mciteSetBstMidEndSepPunct{\mcitedefaultmidpunct}
{\mcitedefaultendpunct}{\mcitedefaultseppunct}\relax
\EndOfBibitem
\bibitem{Forthofer1981}
R.~N. Forthofer and R.~G. Lehnen, in {\em Rank correlation methods}, \href{https://doi.org/10.1007/978-1-4684-6683-6_9}{ pp.~146--163, Springer US, Boston, MA}, 1981\relax
\mciteBstWouldAddEndPuncttrue
\mciteSetBstMidEndSepPunct{\mcitedefaultmidpunct}
{\mcitedefaultendpunct}{\mcitedefaultseppunct}\relax
\EndOfBibitem
\bibitem{COWs}
H.~Dembinski, M.~Kenzie, C.~Langenbruch, and M.~Schmelling, \ifthenelse{\boolean{articletitles}}{\emph{{Custom Orthogonal Weight functions (COWs) for event classification}}, }{}\href{https://doi.org/https://doi.org/10.1016/j.nima.2022.167270}{Nucl.\ Instrum.\ Meth.\  \textbf{A1040} (2022) 167270}, \href{http://arxiv.org/abs/2112.04574}{{\normalfont\ttfamily arXiv:2112.04574}}\relax
\mciteBstWouldAddEndPuncttrue
\mciteSetBstMidEndSepPunct{\mcitedefaultmidpunct}
{\mcitedefaultendpunct}{\mcitedefaultseppunct}\relax
\EndOfBibitem
\bibitem{Xie:2009rka}
Y.~Xie, \ifthenelse{\boolean{articletitles}}{\emph{{sFit: a method for background subtraction in maximum likelihood fit}}, }{}\href{http://arxiv.org/abs/0905.0724}{{\normalfont\ttfamily arXiv:0905.0724}}\relax
\mciteBstWouldAddEndPuncttrue
\mciteSetBstMidEndSepPunct{\mcitedefaultmidpunct}
{\mcitedefaultendpunct}{\mcitedefaultseppunct}\relax
\EndOfBibitem
\bibitem{Zhang2012TimedependentDF}
L.~Zhang and S.~Stone, \ifthenelse{\boolean{articletitles}}{\emph{{Time-dependent Dalitz-plot formalism for $B^0_q \to J/\psi h^+ h^-$}}, }{}\href{https://doi.org/https://doi.org/10.1016/j.physletb.2013.01.035}{Phys.\ Lett.\  \textbf{B719} (2013) 383}, \href{http://arxiv.org/abs/1212.6434}{{\normalfont\ttfamily arXiv:1212.6434}}\relax
\mciteBstWouldAddEndPuncttrue
\mciteSetBstMidEndSepPunct{\mcitedefaultmidpunct}
{\mcitedefaultendpunct}{\mcitedefaultseppunct}\relax
\EndOfBibitem
\bibitem{LHCb-PAPER-2019-013}
LHCb collaboration, R.~Aaij {\em et~al.}, \ifthenelse{\boolean{articletitles}}{\emph{{Updated measurement of time-dependent \CP-violating observables in \mbox{\decay{\Bs}{\jpsi \Kp\Km}} decays}}, }{}\href{https://doi.org/10.1140/epjc/s10052-019-7159-8}{Eur.\ Phys.\ J.\  \textbf{C79} (2019) 706}, Erratum \href{https://doi.org/10.1140/epjc/s10052-020-7875-0}{ibid.\   \textbf{C80} (2020) 601}, \href{http://arxiv.org/abs/1906.08356}{{\normalfont\ttfamily arXiv:1906.08356}}\relax
\mciteBstWouldAddEndPuncttrue
\mciteSetBstMidEndSepPunct{\mcitedefaultmidpunct}
{\mcitedefaultendpunct}{\mcitedefaultseppunct}\relax
\EndOfBibitem
\bibitem{PhysRevLett.87.241801}
BaBar collaboration, B.~Aubert {\em et~al.}, \ifthenelse{\boolean{articletitles}}{\emph{{Measurement of the $\mathit{B}\ensuremath{\rightarrow}\phantom{\rule{0ex}{0ex}}\mathit{J}/\mathit{\ensuremath{\psi}}{\mathit{K}}^{*}(892)$ decay amplitudes}}, }{}\href{https://doi.org/10.1103/PhysRevLett.87.241801}{Phys.\ Rev.\ Lett.\  \textbf{87} (2001) 241801}, \href{http://arxiv.org/abs/hep-ex/0107049}{{\normalfont\ttfamily arXiv:hep-ex/0107049}}\relax
\mciteBstWouldAddEndPuncttrue
\mciteSetBstMidEndSepPunct{\mcitedefaultmidpunct}
{\mcitedefaultendpunct}{\mcitedefaultseppunct}\relax
\EndOfBibitem
\bibitem{BaBar:2004xhu}
{BABAR Collaboration}, B.~Aubert {\em et~al.}, \ifthenelse{\boolean{articletitles}}{\emph{{Time-integrated and time-dependent angular analyses of $B\ensuremath{\rightarrow}J/\ensuremath{\psi}K\ensuremath{\pi}$: A measurement of $\mathrm{cos}﻿2\ensuremath{\beta}$ with no sign ambiguity from strong phases}}, }{}\href{https://doi.org/10.1103/PhysRevD.71.032005}{Phys.\ Rev.\  \textbf{D71} (2005) 032005}, \href{http://arxiv.org/abs/hep-ex/0411016}{{\normalfont\ttfamily arXiv:hep-ex/0411016}}\relax
\mciteBstWouldAddEndPuncttrue
\mciteSetBstMidEndSepPunct{\mcitedefaultmidpunct}
{\mcitedefaultendpunct}{\mcitedefaultseppunct}\relax
\EndOfBibitem
\bibitem{efron:1979}
B.~Efron, \ifthenelse{\boolean{articletitles}}{\emph{Bootstrap methods: Another look at the jackknife}, }{}\href{https://doi.org/10.1214/aos/1176344552}{Ann.\ Statist.\  \textbf{7} (1979) 1}\relax
\mciteBstWouldAddEndPuncttrue
\mciteSetBstMidEndSepPunct{\mcitedefaultmidpunct}
{\mcitedefaultendpunct}{\mcitedefaultseppunct}\relax
\EndOfBibitem
\bibitem{LHCb-PAPER-2016-062}
LHCb collaboration, R.~Aaij {\em et~al.}, \ifthenelse{\boolean{articletitles}}{\emph{{Measurement of \Bd, \Bs, \Bp and \Lb production asymmetries in 7 and 8~\tev\ proton-proton collisions}}, }{}\href{https://doi.org/10.1016/j.physletb.2017.09.023}{Phys.\ Lett.\  \textbf{B774} (2017) 139}, \href{http://arxiv.org/abs/1703.08464}{{\normalfont\ttfamily arXiv:1703.08464}}\relax
\mciteBstWouldAddEndPuncttrue
\mciteSetBstMidEndSepPunct{\mcitedefaultmidpunct}
{\mcitedefaultendpunct}{\mcitedefaultseppunct}\relax
\EndOfBibitem
\bibitem{LHCb-PUB-2018-004}
A.~Davis {\em et~al.}, \ifthenelse{\boolean{articletitles}}{\emph{{Measurement of the instrumental asymmetry for $\Km\pip$-pairs at LHCb in Run 2}}, }{} \href{http://cdsweb.cern.ch/search?p=LHCb-PUB-2018-004&f=reportnumber&action_search=Search&c=LHCb+Notes} {LHCb-PUB-2018-004}, 2018\relax
\mciteBstWouldAddEndPuncttrue
\mciteSetBstMidEndSepPunct{\mcitedefaultmidpunct}
{\mcitedefaultendpunct}{\mcitedefaultseppunct}\relax
\EndOfBibitem
\bibitem{Belle:2002otd}
Belle collaboration, K.~Abe {\em et~al.}, \ifthenelse{\boolean{articletitles}}{\emph{{Measurements of branching fractions and decay amplitudes in $B \to J/\psi K^{*}$ decays}}, }{}\href{https://doi.org/10.1016/S0370-2693(02)01969-X}{Phys.\ Lett.\  \textbf{B538} (2002) 11}, \href{http://arxiv.org/abs/hep-ex/0205021}{{\normalfont\ttfamily arXiv:hep-ex/0205021}}\relax
\mciteBstWouldAddEndPuncttrue
\mciteSetBstMidEndSepPunct{\mcitedefaultmidpunct}
{\mcitedefaultendpunct}{\mcitedefaultseppunct}\relax
\EndOfBibitem
\bibitem{Santos:2013gra}
D.~Mart{\'\i}nez~Santos and F.~Dupertuis, \ifthenelse{\boolean{articletitles}}{\emph{{Mass distributions marginalized over per-event errors}}, }{}\href{https://doi.org/10.1016/j.nima.2014.06.081}{Nucl.\ Instrum.\ Meth.\  \textbf{A764} (2014) 150}, \href{http://arxiv.org/abs/1312.5000}{{\normalfont\ttfamily arXiv:1312.5000}}\relax
\mciteBstWouldAddEndPuncttrue
\mciteSetBstMidEndSepPunct{\mcitedefaultmidpunct}
{\mcitedefaultendpunct}{\mcitedefaultseppunct}\relax
\EndOfBibitem
\bibitem{PhysRevD.90.112009}
Belle collaboration, K.~Chilikin {\em et~al.}, \ifthenelse{\boolean{articletitles}}{\emph{{Observation of a new charged charmoniumlike state in ${\overline{B}}^{0}\ensuremath{\rightarrow}J/\ensuremath{\psi}{K}^{\ensuremath{-}}{\ensuremath{\pi}}^{+}$ decays}}, }{}\href{https://doi.org/10.1103/PhysRevD.90.112009}{Phys.\ Rev.\  \textbf{D90} (2014) 112009}, \href{http://arxiv.org/abs/1408.6457}{{\normalfont\ttfamily arXiv:1408.6457}}\relax
\mciteBstWouldAddEndPuncttrue
\mciteSetBstMidEndSepPunct{\mcitedefaultmidpunct}
{\mcitedefaultendpunct}{\mcitedefaultseppunct}\relax
\EndOfBibitem
\bibitem{Nisius:2020jmf}
R.~Nisius, \ifthenelse{\boolean{articletitles}}{\emph{{BLUE: combining correlated estimates of physics observables within ROOT using the Best Linear Unbiased Estimate method}}, }{}\href{https://doi.org/10.1016/j.softx.2020.100468}{SoftwareX \textbf{11} (2020) 100468}, \href{http://arxiv.org/abs/2001.10310}{{\normalfont\ttfamily arXiv:2001.10310}}\relax
\mciteBstWouldAddEndPuncttrue
\mciteSetBstMidEndSepPunct{\mcitedefaultmidpunct}
{\mcitedefaultendpunct}{\mcitedefaultseppunct}\relax
\EndOfBibitem
\bibitem{fsfd_run1}
LHCb collaboration, R.~Aaij {\em et~al.}, \ifthenelse{\boolean{articletitles}}{\emph{{Measurement of the fragmentation fraction ratio $f_s/f_d$ and its dependence on $B$ meson kinematics}}, }{}\href{https://doi.org/10.1007/JHEP04(2013)001}{JHEP \textbf{04} (2013) 001}, \href{http://arxiv.org/abs/1301.5286}{{\normalfont\ttfamily arXiv:1301.5286}}\relax
\mciteBstWouldAddEndPuncttrue
\mciteSetBstMidEndSepPunct{\mcitedefaultmidpunct}
{\mcitedefaultendpunct}{\mcitedefaultseppunct}\relax
\EndOfBibitem
\end{mcitethebibliography}

\newpage
% LHCb collaboration author list
% Data extracted on June 23rd, 2025 at 10:34am for paper reference LHCb-PAPER-2025-020
\centerline
{\large\bf LHCb collaboration}
\begin
{flushleft}
\small
R.~Aaij$^{38}$\lhcborcid{0000-0003-0533-1952},
A.S.W.~Abdelmotteleb$^{57}$\lhcborcid{0000-0001-7905-0542},
C.~Abellan~Beteta$^{51}$\lhcborcid{0009-0009-0869-6798},
F.~Abudin{\'e}n$^{57}$\lhcborcid{0000-0002-6737-3528},
T.~Ackernley$^{61}$\lhcborcid{0000-0002-5951-3498},
A. A. ~Adefisoye$^{69}$\lhcborcid{0000-0003-2448-1550},
B.~Adeva$^{47}$\lhcborcid{0000-0001-9756-3712},
M.~Adinolfi$^{55}$\lhcborcid{0000-0002-1326-1264},
P.~Adlarson$^{85}$\lhcborcid{0000-0001-6280-3851},
C.~Agapopoulou$^{14}$\lhcborcid{0000-0002-2368-0147}
% ...

{\footnotesize \it

% ...
$^{14}$Universit{\'e} Paris-Saclay, CNRS/IN2P3, IJCLab, Orsay, France\\
% ...
$^{38}$Nikhef National Institute for Subatomic Physics, Amsterdam, Netherlands\\
% ...
$^{47}$Instituto Galego de F{\'\i}sica de Altas Enerx{\'\i}as (IGFAE), Universidade de Santiago de Compostela, Santiago de Compostela, Spain\\
% ...
$^{51}$Physik-Institut, Universit{\"a}t Z{\"u}rich, Z{\"u}rich, Switzerland\\
% ...
$^{55}$H.H. Wills Physics Laboratory, University of Bristol, Bristol, United Kingdom\\
% ...
$^{57}$Department of Physics, University of Warwick, Coventry, United Kingdom\\
% ...
$^{60}$School of Physics and Astronomy, University of Glasgow, Glasgow, United Kingdom\\
$^{61}$Oliver Lodge Laboratory, University of Liverpool, Liverpool, United Kingdom\\
% ...
$^{69}$Syracuse University, Syracuse, NY, United States\\
% ...
$^{85}$Department of Physics and Astronomy, Uppsala University, Uppsala, Sweden, associated to $^{60}$\\
% ...
}
\end{flushleft}
% \input{Authorship_LHCb-PAPER-2025-020-arXiv}
%\input{~/Downloads/Authorship_LHCb-SD-2024-001-grouped-internal-use-only-do-not-publish-2.tex}

% The author list for journal publications is generated from the
% Membership Database shortly after 'approval to go to paper' has been
% given.  It is available at \url{https://lbfence.cern.ch/membership/authorship}
% and will be sent to you by email shortly after a paper number
% has been assigned.  
% The author list should be included in the draft used for 
% first and second circulation, to allow new members of the collaboration to verify
% that they have been included correctly. Occasionally a misspelled
% name is corrected, or associated institutions become full members.
% Therefore an updated author list will be sent to you after the final
% EB review of the paper.  In case line numbering doesn't work well
% after including the authorlist, try moving the \verb!\bigskip! after
% the last author to a separate line.

% The authorship for Conference Reports should be ``The LHCb
% collaboration'', with a footnote giving the name(s) of the contact
% author(s), but without the full list of collaboration names.

% The authorship for Figure Reports should be ``The LHCb
% collaboration'', with no contact author and without the full list 
% of collaboration names.

\end{document}